\documentclass[conference]{IEEEtran}
\usepackage{tikz}
\usepackage{amsmath}

\usepackage[english]{babel}
\usepackage{graphicx}
\usepackage{graphics}
\usepackage{caption,subcaption}
\usepackage{amsmath}
\usepackage[cmintegrals]{newtxmath}
\usepackage{url}
\usepackage{listings}
\usepackage[square, numbers]{natbib}
\usepackage[parfill]{parskip}
\usepackage{xcolor}
\usepackage{xspace}
\usepackage{booktabs}
\usepackage{tablefootnote}
\usepackage{colortbl}
\usepackage{xintexpr, xinttools}

\newcommand{\cTab}[1]{
\cellcolor{red!\xinttheiexpr 70 * #1 * #1\relax}
\xintifboolexpr {#1 > 0.8 * 1}{\textcolor{white}{#1}}{#1} 
}
\newcommand{\para}[1]{\smallskip \noindent \textbf{#1}}
\newcommand{\parait}[1]{\smallskip \noindent \textit{#1}}

\newcommand\eg{\emph{e.g.},\xspace}
\newcommand\ie{\emph{i.e.},\xspace}

\providecommand{\etal}{\emph{et al.}\xspace}

\newif\ifvarcomm

\varcommfalse
\ifvarcomm

\else

\fi

\def \loca {LOC1\xspace} % Lausanne
\def \locb {LOC2\xspace} % Leuven
\def \locc {LOC3\xspace} % Singapore
\def \ff {CL-FF\xspace}
\def \clia {\cloud}% cl: cloudflare client
\def \clib {\ff}   % cc/ff: firefox + trr
\def \clic {\locb} % ch: (chrome + cloduflare client)
\def \desktop {DESKTOP\xspace}
\def \rpi {RPI\xspace}
\def \google {GOOGLE\xspace}
\def \cloud {CLOUD\xspace}
\def \tor {TOR\xspace}
\def \ow {OW\xspace}
\def \wfp {WEB\xspace}
\def \ednscloud {EDNS0-128\xspace}
\def \ednsad {EDNS0-128-adblock\xspace}
\def \ednsrec {EDNS0-468\xspace}
\def \dot {DOT\xspace}

\def\ffpages{700\xspace}
\def\ffsamples{60\xspace}
\def \owpages {5,000\xspace}
\def \cwpages {1,500\xspace}
\def \cwpagesmini {700\xspace}
\def \owsamples {3\xspace}
\def \cwsamplesmax {200\xspace}
\def \cwsamples {60\xspace}

\begin{document}
\date{}

\title{\Large \bf Encrypted DNS $\implies$ Privacy? A Traffic Analysis Perspective}

%\author{\IEEEauthorblockN{Anonymous}
\author{\IEEEauthorblockN{Sandra Siby}
\IEEEauthorblockA{EPFL}
\and
\IEEEauthorblockN{Marc Juarez}
\IEEEauthorblockA{imec-COSIC KU Leuven}
\and
\IEEEauthorblockN{Claudia Diaz}
\IEEEauthorblockA{imec-COSIC KU Leuven }
\and
\IEEEauthorblockN{Narseo Vallina-Rodriguez}
\IEEEauthorblockA{IMDEA Networks }
\and
\IEEEauthorblockN{Carmela Troncoso}
\IEEEauthorblockA{EPFL}
}

\maketitle

%!TEX root = report.tex
\begin{abstract}
Virtually every connection to an Internet service is preceded by a DNS lookup
which is  
performed without any traffic-level protection, thus enabling manipulation, 
redirection, surveillance, and censorship. 
To address these issues, large organizations such as Google and Cloudflare are deploying 
recently standardized protocols that encrypt DNS traffic between
end users and recursive resolvers such as DNS-over-TLS (DoT) and DNS-over-HTTPS (DoH).
In this paper, we examine whether encrypting DNS traffic can protect users
from traffic analysis-based monitoring and censoring. 
We propose a novel feature set to perform the attacks, as those used to attack HTTPS or Tor traffic
are not suitable for DNS' characteristics. We show that traffic analysis enables the identification of domains 
with high accuracy in closed and open world settings,
using 124 times less data than attacks on HTTPS flows.
We find that factors such as location, resolver, platform, or client do mitigate the attacks performance but they are far from completely stopping them.
Our results indicate that DNS-based censorship is still possible on encrypted DNS traffic. 
In fact, we demonstrate that the standardized padding schemes are not effective. Yet, 
Tor ---which does not effectively mitigate
traffic analysis attacks on web traffic--- is a good defense against DoH traffic analysis.

\end{abstract}

\section{Introduction}
\label{sec:introduction}
%!TEX root = report.tex
Regular Domain Name System (DNS) requests have been mostly sent in the clear~\cite{bortzmeyer2015dns}.
This situation enables entities, such as Internet Service Providers (ISPs), Autonomous Systems (ASes), 
or state-level agencies, to perform user tracking, mass surveillance~\cite{GrothoffWEA17,thensafiles}
and censorship~\cite{dns-turkey,weaver2011redirecting}.
The risk of pervasive surveillance and its consequences has prompted 
Internet governance, industry actors, and standardization bodies to foster 
privacy protections~\cite{rfc7626,rfc6973}. 
In particular, for DNS, these bodies have standardized two protocols:
DNS-over-TLS (DoT)~\cite{dot} and DNS-over-HTTPS (DoH)~\cite{doh}. These
protocols encrypt the communication between the client and the recursive resolver
to prevent the inspection of domain names
by network eavesdroppers. These standarization bodies also consider protection
mechanisms to limit inference of private information from traffic metadata,
such as the timing and size of network packets, of the encrypted DNS
communication.

These mechanisms protect against traffic analysis by padding traffic 	
\cite{edns0padding}, or by multiplexing the encrypted DNS traffic with
other traffic, \eg when DoH and web HTTPS traffic share a single TLS
tunnel (see \S8.1~\cite{doh}). 

During 2018, Google and Cloudflare
launched public DoH resolvers~\cite{Google, Cloudflare}, while Mozilla 
added DoH support to Firefox~\cite{ff-doh}, which will become the 
default DNS protocol version for US users in September 2019~\cite{FirefoxDefault}. 
These efforts aim to leverage DoH and DoT's 
capacity to enhance users' browser traffic security guarantees~\cite{doh-explained}. 
Yet, it is known that even when communications are encrypted,
traffic features such as volume and timing can reveal 
information about their content~\cite{PanchenkoLZHPWE16, WangG16, HayesD16,
SirinamIJW18,WhiteMSM11,OhLH17}, and whether 
measures like padding are sufficient to protect users against eavesdroppers. 
As of today, existing evaluations of DoH implementations have focused
on understanding the impact of encryption and transport protocols
on performance~\cite{dns-perf-measQ1-18, dns-perf-measQ4-18}, 
and on cost~\cite{HounselBSHF19,BottgerCAFTCU19}.

In this paper, we aim to fill this gap by studying the effectiveness of traffic analysis attacks
in revealing users' browsing patterns from encrypted DNS.
We focus our analysis on DoH, as its adoption by large industry actors (e.g., Google, Cloudflare, and Mozilla) makes it prevalent in the wild. 
For completeness, we include a comparison of the protection provided by 
Cloudflare's DoH and DoT resolvers.
In our analysis, we consider a passive adversary, as described in RFC
7626~\cite{rfc7626}, who is placed between the client and the 
DNS resolver. The adversary's goal is to identify which web pages users visit, 
to either perform surveillance or censorship.
As the RFC stresses, this adversary may be in ``a different path than the communication between the initiator [\eg the client] and the
recipient [\eg a web server]''~\cite{rfc7626}, and thus can launch attacks even if they do not see
all the traffic between endpoints. 

We find that features traditionally used in atttacks on web
traffic~\cite{PanchenkoLZHPWE16, WangG16,HayesD16, SirinamIJW18, MillerHJT14,
LuoZCLCP11} are not suitable for encrypted DNS traffic. As opposed to web
traffic, DNS traffic ---even if encrypted---
is bursty, chatty, and is mostly composed of small packets.
We engineer a \emph{novel set of features} that, by focusing on local traffic features,
enables successful attacks that identify requested websites on encrypted DNS.
As encrypted DNS traces are much smaller that their web traffic counterparts, our 
techniques require \emph{124 times less data than state-of-the-art traffic analysis on web traffic},
which is key to ensure scalability of attacks~\cite{NasrHM17}. 
Furthermore, our new feature set on encrypted DNS traffic is \emph{as effective or more
so} than state-of-the-art attacks on web traffic in identifying web pages.

We also find that differences between the environment used by the adversary to
train the attack (e.g., location, choice of client application, platform or
recursive DNS resolver), and the environment where the attack is actually
deployed, negatively affect the performance of the attack.  Most prior work on
traffic analysis assumes the adversary knows the environment where the attack
will be deployed, but the adversary cannot trivially obtain that information a
priori. Our features allow the adversary to infer that information and thus tailor the
attacks accordingly, maintaining high attack performance for each specific environment.

Next, we evaluate existing traffic analysis defenses,
including the standardized EDNS0 padding~\cite{edns0padding} ---implemented 
by Cloudflare and Google in their solutions---,
and the use of Tor~\cite{dotor} as transport, a feature 
available when using Cloudflare's resolver. We find that,
unless EDNS0 padding overhead is large, current padding strategies cannot completely prevent our attack. 
Also, while Tor offers little protection against web page fingerprinting on web traffic~\cite{WangG16, HayesD16, SirinamIJW18,WangG13},
Tor is an extremely effective defense against web page fingerprinting on
encrypted DNS traffic.

Finally, we measure the potential of encryption to hinder  
DNS-based censorship practices. We show that under encryption,
it is still possible to identify which packet carries the DNS lookup for the
first domain. We quantify the collateral damage of blocking the response to this
lookup, thereby preventing the user from seeing any content. We also show that, to minimize the effect of 
censorship on non-blacklisted websites, censors must wait to see, on average,
15\% of the encrypted DNS traces.

Our main contributions are as follows:

\begin{itemize}%[noitemsep,nolistsep,leftmargin=11pt]
\item We show that the features for traffic analysis existing
in the literature are not effective on encrypted DNS. We propose a new feature set 
that results in successful attacks on encrypted DNS and that outperforms
existing attacks on HTTPS (Section~\ref{sec:classifier}).  
\item We show that web page fingerprinting on DoH
achieves the same accuracy as web page fingerprinting on encrypted web traffic, while requiring \emph{124} times less volume of data. 
We show that factors such as end-user location, choice of  
DNS resolver, and client-side application or platform, have a negative impact on the effectiveness of the attacks
but do not prevent them (Section~\ref{sec:fingerprintability}).
\item We evaluate the defenses against traffic analysis proposed in the standard and
show that they are not effective. We find that in the case of encrypted DNS,
contrary to web traffic, routing over Tor deters web page identification on encrypted DNS traffic (Section~\ref{sec:countermeasures}).
\item We evaluate the feasibility of DNS-based censorship when DNS lookups are encrypted. We show
that the censor can identify the packet with the first domain lookup. We quantify the tradeoff 
between how much content from a blacklisted site the user can download, and how many non-blacklisted websites are censored
as a side effect of traffic-analysis-based blocking (Section~\ref{sec:censorship}).
\item We gather the first dataset of encrypted DNS traffic collected
in a wide range of environments (Section~\ref{sec:collection}).\footnote{Our dataset and code will be made public upon acceptance.}
\end{itemize}

\noindent\textbf{Impact}
Upon responsible disclosure
of our attacks, Cloudflare changed their DoH resolver to include
padding. This work was also invited to an IETF Privacy Enhancements and Assessments 
Research Group Meeting and will contribute to the next RFC for traffic analysis
protection of encrypted DNS.

\section{Background and Related Work}
\label{sec:background}
%!TEX root = report.tex
In this section, we provide background on the Domain Name System (DNS) and 
existing work on DNS privacy. 

\para{The Domain Name System (DNS)}
is primarily used for translating easy-to-read domain names to numerical IP
addresses~\footnote{Over time, other applications have been 
built on top of DNS~\cite{iodine,spamhaus}}. 
This translation is known as domain resolution.
In order to resolve a domain, a client sends a DNS query to a
\emph{recursive resolver}, a server typically provided by the ISP 
with resolving and caching capabilities. 
If the domain resolution by a client is not cached by the recursive name server,
it contacts a number of \emph{authoritative name servers} 
which hold a distributed database of domain names to IP mappings. 
The recursive resolver traverses the hierarchy of authoritative name servers until it obtains
an answer for the query, and sends it back to the client. The client
can use the resolved IP address to connect to the destination host. Figure~\ref{fig:overview}
illustrates this process.

\para{Enhancing DNS Privacy.}
Security was not a major consideration in the first
versions of DNS, and for years DNS traffic was sent in the
clear over (untrusted) networks. Over the last few years, 
security and privacy concerns have fostered 
the appearance of solutions to make DNS traffic 
resistant to eavesdropping and tampering. Several studies have empirically
demonstrated how the open nature of DNS traffic is being abused for
performing censorship~\cite{Anonymous14, pearce2017global} and surveillance~\cite{GuhaF07,GrothoffWEA17}.
Early efforts include protocols such as
DNSSEC~\cite{dnssec} and DNSCrypt~\cite{dnscrypt}. 
DNSSEC prevents manipulation of DNS data using digital signatures. 
It does not, however, provide confidentiality.
DNSCrypt, an open-source effort, provides both confidentiality
and authenticity. However, due to lack of standardization, it has not achieved
wide adoption. 

In 2016, the IETF approved DNS-over-TLS (DoT)~\cite{dot} as a Standards
Track protocol. The client establishes a TLS session with a recursive resolver (usually on port TCP:853 as
standardized by IANA~\cite{dot}) and exchanges DNS queries and responses over the encrypted connection.
To amortize costs, the TLS session between the client and the recursive
DNS resolver is usually kept alive and reused for multiple queries. Queries go through this
channel in the same manner as in unencrypted DNS -- chatty and in small volume. 

In DoH, standardized in 2018, the local DNS resolver establishes an HTTPS connection to the 
recursive resolver and encodes the DNS queries in the body of HTTP requests. 
DoH considers the use of HTTP/2's Server Push mechanism.
This enables the server to preemptively push DNS responses that
are likely to follow a DNS lookup~\cite{dotvsdoh}, thus 
reducing communication latency.
As opposed to DoT, which uses a dedicated TCP port
for DNS traffic and thus it is easy to monitor and block, DoH lookups
can be sent along non-DNS traffic using existing HTTPS connections (yet
potentially blockable at the IP level).

There are several available implementations of DoT and DoH.
Since 2018, Cloudflare and Quad9 provide both DoH and DoT resolvers, Google
supports DoH, and Android Pie has native support for DoT.
DoH enjoys widespread support from 
browser vendors. Firefox provides the option of directing DNS
traffic to a \emph{trusted recursive resolver} such as a DoH resolver, 
falling back to plaintext DNS if the resolution over DoH fails.
In September 2019, Google announced support for DoH in version 78 of Chrome~\cite{ch-doh}.
Cloudflare also distributes a stand-alone DoH client and, in 2018, they released a hidden resolver that provides DNS over Tor,
not only protecting lookups from eavesdroppers but also providing anonymity 
for clients towards the resolver. Other protocols, 
such as DNS-over-DTLS~\cite{dodtls}, an Experimental RFC proposed by Cisco in 2017,
and DNS-over-QUIC~\cite{doquic}, proposed to the IETF in 2017 by industry 
actors, are not widely deployed so far.

Several academic works study privacy issues related to encrypted DNS.
Shulman suggests that encryption alone may not be sufficient to protect 
users~\cite{shulman2014pretty} but does not provide any
experiments that validate this statement. Our results confirm her hypothesis that
encrypted DNS response size variations can be a distinguishing feature. 
Herrmann et al.\ study the potential of DNS traces as identifiers to perform 
user tracking but do not consider encryption~\cite{herrmann2013behavior}.
Finally, Imana et al.\ study privacy leaks on traffic between 
recursive and authoritative resolvers~\cite{imana2017enumerating}. This is not protected by DoH and 
it is out of scope of our study.

\section{Problem Statement}
\label{sec:goals}
%!TEX root = report.tex

In this paper, we study if it is possible to infer which 
websites a user visits by observing encrypted DNS traffic.
This information is of interest to multiple actors, \eg entities computing statistics on Internet
usage~\cite{dns-oarc,dns-icann}, entities looking to identify malicious
activities~\cite{dns-zero,dns-analytics,weaver2011redirecting}, entities performing 
surveillance~\cite{GuhaF07,GrothoffWEA17}, or entities
conducting censorship~\cite{Anonymous12,pearce2017global}.

\begin{figure}[t!]
\centering
\includegraphics[width=1\linewidth]{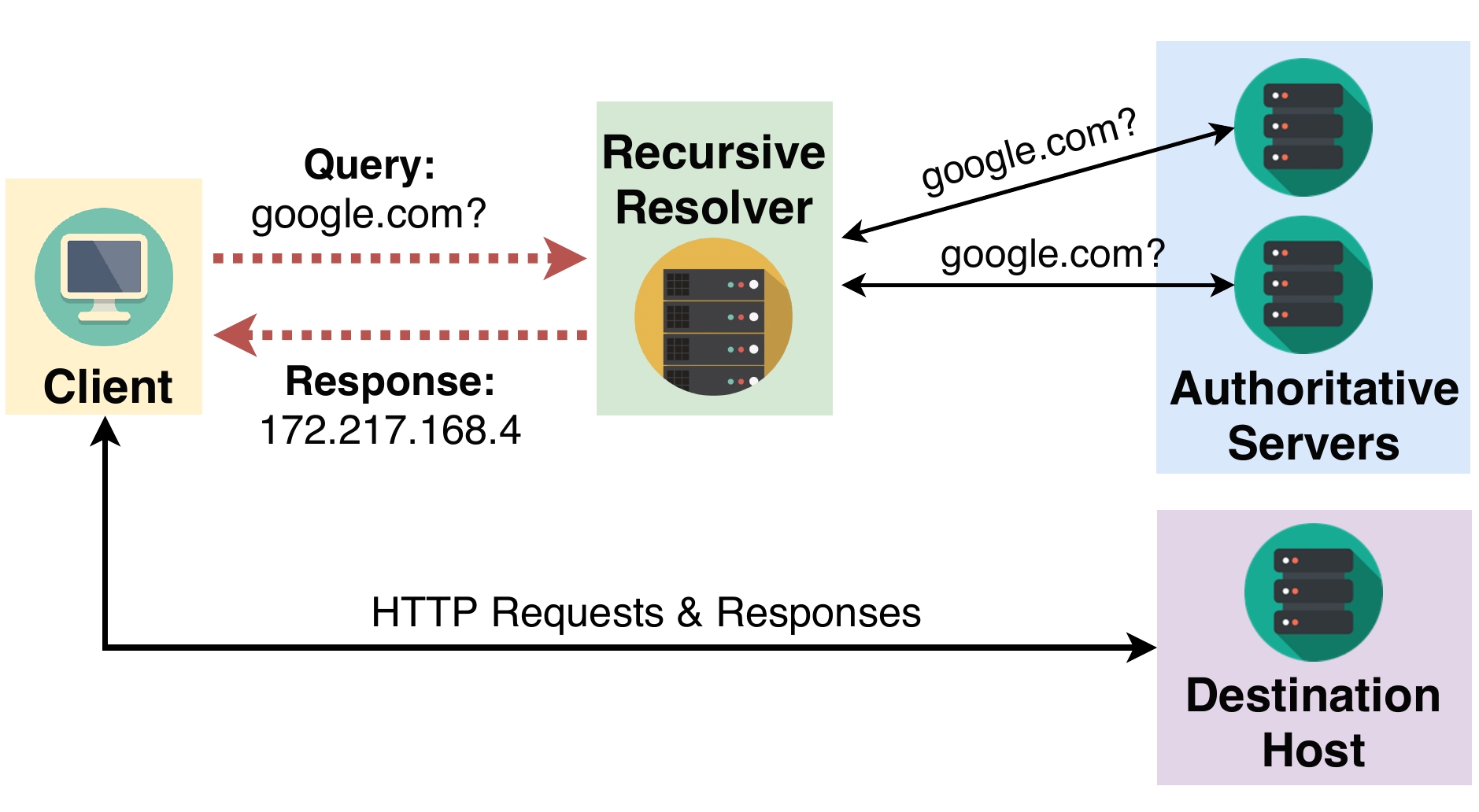}
	\caption{DNS resolution: To visit
\texttt{www.google.com}, a user queries the recursive resolver for its IP. If the
record is not cached, the recursive
	resolver queries an authoritative resolver and forwards the response to
the client. The client uses the IP in the response to connect to the server via HTTP.
	We consider an adversary placed between the client and the resolver (i.e., observes the red dotted lines).}
\label{fig:overview}
	\vspace{-0.7cm}
\end{figure}

We consider an adversary that can collect encrypted DNS traffic between the user and 
the DNS recursive resolver (red dotted lines in Figure~\ref{fig:overview}),
and thus, can link lookups to a specific origin IP address. 
Such an adversary could be present in the users' local network, 
near the resolver, or anywhere along the path (e.g., an ISP or compromised network router). 
As noted in the RFC, this adversary may be ``in a
different path than the communication between the initiator and the
recipient''~\cite{rfc7626}.

This adversary model also includes adversaries that only see DoH traffic, e.g.,
adversaries located in an AS that lies between the user and the resolver but not between the user and the destination host.
Measurements performed from our university network show that this is the case in a significant number of cases. Furthermore,
we note that BGP hijacking attacks, which are becoming increasingly 
frequent~\cite{bgpstream}, can be used to selectively intercept paths to DoH resolvers.
In such cases, the adversary can only rely on DNS fingerprinting 
to learn which webpages are visited by a concrete user for monitoring,
or censorship~\cite{GuhaF07,GrothoffWEA17}. 
If the adversary is actually in the path to the users' destination, she could perform traditional website fingerprinting.
However, we show that web page fingerprinting on only DoH traffic achieves the same accuracy while requiring (on average)
124 times less data than attacking HTTPS traces that also include web traffic. This is critical to
guarantee the scalability of attacks to a large number of targets~\cite{NasrHM17}.

We assume that the adversary has access to \emph{encrypted DNS traffic traces} that 
are generated when the user visits a website via HTTP/S using DoH to
resolve the IPs of the resources. 
A encrypted DNS trace, which we also call DoH trace, comprises the resolution of 
the visited website's first-party domain, and 
the subsequent resolutions for the resources contained in the website, e.g.,
images and scripts. For instance, for visiting Reddit, after resolving
\texttt{www.reddit.com}, the client
would resolve domains such as
\texttt{cdn.taboola.com}, \texttt{doubleclick.net} and
\texttt{thumbs.redditmedia.com}, among others. 

We consider two adversarial goals. First, \emph{monitoring}
the browsing behavior of users, which we study in Section~\ref{sec:fingerprintability};
and, second, \emph{censoring} the web pages that users visit, which we address in Section~\ref{sec:censorship}.
These two goals differ in their data collection needs. Monitoring adversaries can collect full traces to 
make their inferences as accurate as possible, as they do not take any action based on their observations.
In contrast, censorship adversaries need to find out which
domain is being requested as fast as possible so as to interrupt the
communication. Thus, they act on partial traces.

\section{Data Collection}
\label{sec:collection}
%!TEX root = report.tex

To collect data we set up Ubuntu 16.04 virtual machines with  
DoH clients that send DNS queries to a public DoH resolver. 
We use Selenium\footnote{https://www.seleniumhq.org/} (version 3.14.1) to
automatically visit a webpage from our list, triggering DNS lookups for its resources. 
We restart the browser in between webpage visits to ensure that the cache and profile do not
affect collection. We capture network traffic between the DoH client and the resolver using \textit{tcpdump},
and we filter the traffic by destination port and IP to obtain the final DoH traffic trace.

We collect traces for the top, middle, and bottom 500 webpages in Alexa's top 
million websites list on 26 March 2018, \cwpages webpages in total.
We note that even though 1,500 is still a small world 
compared to the size of the Web, the largest world considered in evaluations
similar to ours for website fingerprinting on web traffic over Tor is of 800
websites~\cite{CaiZJJ12,WangG13,WangCNJG14,HayesD16,RimmerPJVW18,SirinamIJW18}.

We visit each webpage in a round-robin fashion, 
obtaining up to \cwsamplesmax samples for every webpage. 
For our open world analysis, we collect traces of an additional \owpages
webpages from the top domains of the Alexa list.
We collected data during two periods, from 26 August 2018 to 9 November 2018,
and from 20 April 2019 to 14 May 2019. Data from these two periods
is never mixed in the analysis.

We consider different scenarios varying DoH 
client and resolver, end user location and platform,
and the use of DNS traffic analysis defenses. 
Table~\ref{tab:datasets} provides an overview of the collected datasets.
In order to better understand the vulnerability of DNS encryption to traffic analysis,
we designed heterogenous experiments that look into different aspects of the
problem, resulting in multiple datasets of varied sizes and collected under
different conditions  -- in many cases, several datasets for each experiment type.
We also collected
a dataset (using the Stubby client and CLoudflare's resolver) to study the effectiveness of defenses on
DoT traffic as compared to DoH. Since Cloudflare had already implemented padding of responses for DoT
traffic at the time of data collection, we were not able to collect a dataset of
DoT without padding.
In the following sections, we use the Identifier provided in the
second column to refer to each of the datasets. 
Note that unless specified otherwise, we use Cloudflare's DoH client.

\begin{table}[t!]
  \centering
  \footnotesize
  \caption{Overview of datasets.}
  \resizebox*{\columnwidth}{!}{%
  \begin{tabular}[c]{ c c c c}
  \toprule
    \textbf{Name} & \textbf{Identifier} &  \textbf{\# webpages} & \textbf{\# samples}\\
    \midrule
    Desktop (Location 1) & \loca & \cwpages & \cwsamplesmax\\
    Desktop (Location 2) & \locb & \cwpages & \cwsamples \\
    Desktop (Location 3) & \locc & \cwpages & \cwsamples \\
    Raspberry Pi & \rpi & \cwpagesmini & \cwsamples \\
    Firefox with Google resolver & \google & \cwpagesmini & \cwsamples \\
    Firefox with Cloudflare resolver & \cloud & \cwpagesmini & \cwsamples \\ 
    Firefox with Cloudflare client & \ff & \ffpages & \ffsamples\\
    Open World & \ow & \owpages & \owsamples \\
    DoH and web traffic & \wfp & \cwpagesmini & \cwsamples \\
    DNS over Tor & \tor & \cwpagesmini & \cwsamples \\
    Cloudflare's EDNS0 padding implementation  & \ednscloud & \cwpagesmini & \cwsamples \\
    Recommended EDNS0 padding & \ednsrec & \cwpagesmini & \cwsamples \\
    EDNS0 padding with ad-blocker & \ednsad & \cwpagesmini & \cwsamples \\
    DoT with Stubby client & \dot & \cwpagesmini & \cwsamples \\
    \bottomrule
  \end{tabular}}
  \label{tab:datasets}
\end{table}

\para{Data curation.} 
We curate the datasets to ensure that 
our results are not biased. First, we aim at removing the effect of spurious errors in 
collection. We define those as changes in the websites for
reasons other than those variations due to their organic evolution
that do not represent the expected behavior of the page. For instance, 
pages that go down during the collection period.

Second, we try to eliminate website behaviors 
that are bound to generate classification 
errors unrelated to the characteristics of DNS traffic. 
These occur when different domains generating the same exact DNS traces,
e.g., when webpages redirect to other pages or to 
the same resource, or when web servers return the same errors 
(e.g., 404 not found or 403 forbidden).

We use the Chrome over Selenium crawler to collect 
the HTTP request/responses, not the DNS queries responses, of all the 
pages in our list in \loca.
We conduct two checks. First, we look at the HTTP
response status of the \textit{top level domain}, i.e., the URL
that is being requested by the client. We identify the webpages that do not have an
HTTP OK status. These could be caused by a number of factors, such as pages
not found (404), anti-bot mechanisms, 
forbidden responses due to geoblocking~\cite{mcdonald2018403} (403), 
internal server errors (500), and so on.
We mark these domains as conflicting.
Second, we confirm that the top level domain is
present in the list of requests and responses. This ensures that the page
the client is requesting is not redirecting the browser to other URLs. 
This check triggers some false alarms. For example, a webpage might redirect to a
country-specific version (\texttt{indeed.com} redirecting to
\texttt{indeed.fr}, results in \texttt{indeed.com} not being present in the
list of requests); or in domain redirections (\texttt{amazonaws.com}
redirecting to \texttt{aws.amazon.com}). We do not consider these cases as anomalies.
Other cases are full redirections. Examples are malware that redirect 
browser requests to \texttt{google.com},
webpages that redirect to GDPR country restriction notices, or
webpages that redirect to domains that specify that the site is closed. We consider these
cases as invalid webpages and add them to our list of conflicting domains.

We repeat these checks multiple times over our collection period. 
We find that 70 webpages that had invalid statuses at some point during
our crawl, and 16 that showed some fluctuation in their status 
(from conflicting to non-conflicting or vice versa). We
study the effects of keeping and removing these conflicting webpages in
Section~\ref{sec:ngrams}.

\section{DNS-based Website Fingerprinting}
\label{sec:fingerprintability}
%!TEX root = report.tex
Website fingerprinting attacks enable a local
eavesdropper to determine which web pages a user is accessing
over an encrypted or anonyimized channel. It exploits the fact that the size, timing,
and order of TLS packets are a reflection of a website's content. 
As resources are unique to each webpage, the traces identify the web even if traffic is encrypted. 
Website fingerprinting has been
shown to be effective on HTTPS~\cite{LuoZCLCP11, MillerHJT14,gonzalez2016user}, 
OpenSSH tunnels~\cite{LiberatoreLL06,DyerCRS12}, encrypted web 
proxies~\cite{SunSWRPQ02, Hintz03} and VPNs~\cite{HerrmannWF09}, and even on
anonymous communications systems such as Tor~\cite{PanchenkoNZE11, WangG13,
CaiZJJ12, WangCNJG14, PanchenkoLZHPWE16, WangG16, HayesD16, SirinamIJW18}.

The patterns exploited by website fingerprinting 
are correlated with patterns in DNS traffic:  
which resources are loaded and their order determines 
the order of the corresponding DNS queries. 
Thus, we expect that website fingerprinting
can also be done on DNS encrypted flows  
such as DNS-over-HTTPS (DoH) or DNS-over-TLS (DoT).
In this paper, we call \emph{DNS fingerprinting} the use of traffic 
analysis to identify the web page that generated an encrypted
DNS trace, i.e., website fingerprinting on encrypted DNS traffic. 
In the following, whenever we do 
not explicitly specify whether the target of website fingerprinting is DNS or HTTPS traffic, 
we refer to traditional website fingerprinting on web traffic over HTTPS.

\subsection{DNS fingerprinting}\label{sec:classifier}
As in website fingerprinting, we treat DNS fingerprinting as a 
supervised learning problem: the adversary first collects a training dataset
of encrypted DNS traces for a set of pages. The page (label)
corresponding to a DNS trace is known. The adversary extracts
features from these traces (e.g., lengths of network packets) and
trains a classifier that, given a network trace, identifies the visited page. 
Under deployment the adversary collects traffic from a victim and feeds it to the classifier
to determine which page the user is visiting.

\para{Traffic variability.}
Environmental conditions introduce variance in the DNS traces sampled for the same website. 
Thus, the adversary must collect multiple DNS traces in order to 
obtain a robust representation of the page. 

Some of this variability has similar origin to that of web traffic. For instance,
changes on the website that result on varying
DNS lookups associated with third-party embedded resources (e.g., ads shown to users);
the platform where the client runs, the configuration
of the DoH client, or the software using the client which may vary the DNS requests
(e.g., mobile versions of websites, or browsers' use of pre-fetching);
and the effects of content localization and personalization that determine which 
resources are served to the user.

Additionally, there are some variability factors specific to DNS traffic.
For instance, the effect of the local resolver, which depending on the state of 
the cache may or may not launch requests to the authoritative server; 
or the DNS-level load-balancing and replica selection (e.g., CDNs) 
which may provide different IPs for a given domain or resource~\cite{almeida2017dissecting}.

\para{Feature engineering.}
Besides the extra traffic variability compared to web traffic, 
DNS responses are generally smaller and chatty than web
resources~\cite{almeida2017dissecting,plonka2008context}.
In fact, even when DNS lookups are wrapped in HTTP requests, DNS requests and responses fit in one single TLS record
in most cases. These particularities hint that traditional website fingerprinting features,
typically based on aggregate metrics of traffic traces such as the total number of packets,
total bytes, and their statistics (e.g., average, standard deviation), are inadequate to
characterize DoH traffic. We test this hypothesis in the following section, where
we show that state-of-the-art web traffic attacks' performance drops in 20\%
when applied on DoH traces (see Table~\ref{tab:webfp_analysis}). 

To address this problem, we introduce a novel set of features,
consisting of n-grams of TLS record lengths in a trace. Following
the usual convention in website fingerprinting, we represent a traffic trace as
a sequence of integers, where the absolute value is the size of the TLS record
and the sign indicates direction: positive for packets from the client to the
resolver (outgoing),
and negative for the packets from the resolver to the client (incoming). An
example of this represenation is the trace: $(-64, 88, 33, -33)$.
Then, the uni-grams for this trace are $(-64), (88), (33), (-33)$, and the bi-grams are $(-64, 88),
(88, 33), (33, -33)$.  
To create the features, we take tuples of $n$ consecutive TLS
record lengths in the DoH traffic traces and count the number
of their occurrences in each trace.
The intuition behind our choice is that n-grams capture patterns in 
request-response size pairs, as well as the local 
order of the packets sequence. 
To the best of our knowledge, n-grams have never been considered as features in the website 
fingerprinting literature.
 
We extend the n-gram representation to traffic bursts.
Bursts are sequences of consecutive packets in the same direction
(either incoming or outgoing). Bursts correlate with the number and order of
resources embedded in the page. Additionally, they are more robust to small changes in
order than individual sizes because they aggregate several records in the same
direction. We represent n-grams of bursts by adding lengths on packets in 
a direction inside the tuple. In the previous example, the burst-length sequence of the
trace above is $(-64, 121, -33)$ and the burst bi-grams are $(-64, 121), (121,
-33)$.

We experimented with uni-, bi- and tri-grams for both features types. We
observed a marginal improvement in the classifier on using tri-grams at a
substantial cost on the memory requirements of the classifier.
%, since the space of feature explodes when the size of the n-grams increases. 
We also experimented with the timing of packets. As in website
fingerprinting~\cite{WangG13}, we found that it does not provide 
reliable prediction power. This is because timing greatly varies depending on the state of the 
network and thus is not a stable feature to fingerprint web pages.
In our experiments, we use the concatenation of uni-grams and
bi-grams of both TLS record sizes and bursts as feature set.

\para{Algorithm selection.} After experimenting with different supervised
classification algorithms, we selected Random Forests (RF)
which are known to be very effective for traffic analysis
tasks~\cite{HayesD16, JansenJGED18}. 

Random forests (RF) are ensembles of simpler classifiers called decision trees.
Decision trees use a tree data
structure to represent splits of the data: nodes represent a condition on one
of the data features and branches represent decisions based on the evaluation
of that condition. 
In decision trees, feature importance in classification is measured with respect to
how well they split samples with respect to the target classes. The more skewed 
the distribution of samples into classes is, the better the feature discriminates. Thus,
a common metric for importance is the Shannon's entropy of this distribution. 
Decision trees, however, do not generalize well and tend to overfit the
training data. RFs mitigate this issue by randomizing the data and features
over a large amount of trees, using different subsets of features and data in
each tree. The final decision of the RF is an aggregate function on
the individual decisions of its trees. In our experiments, we use 100
trees and a majority vote to aggregate them.
%\narseo{Citations might be good in the paragraph above}

\para{Validation.} We evaluate the effectiveness of our classifier measuring
the \emph{Precision}, \emph{Recall} and \emph{F1-Score}. 
Considering positives as websites correctly identified, and negatives websites
classified to an incorrect label: Precision is the ratio of true
positives to the total number of samples that were classified
as positive (true positives and false positives); Recall is the
ratio of true positives to the total number of positives (true
positives and false negatives); and the F1-score is the harmonic
mean of Precision and Recall. 

We evaluate the effectiveness of DNS fingerprinting in two scenarios 
typically used in the web traffic analysis literature. 
A \emph{closed world}, in which the adversary knows the set of all
possible webpages that users may visit; and an \emph{open-world}, in which the
adversary only has access to a set of \emph{monitored} sites, and the user may
visit webpages outside of this set.

We use 10-fold cross-validation, a standard methodology in machine learning, to eliminate biases
due to overfitting in our experiments. In cross-validation,
the samples of each class are divided in ten disjoint sets. The classifier is
then trained on nine of the sets and tested in the remaining one,
proving ten samples of the classifier performance on a set of samples on which 
it has not been trained on. This estimates the performance of the classifier to unseen examples.

% For the open-world, we have adapted previous cross-validation following previous
% open-world studies, as detailed in Section~\ref{sec:open-world}.

\subsection{Evaluating $n$-grams features} \label{sec:ngrams}

We now evaluate the effectiveness of n-grams features to launch DNS fingerprinting attacks.
We also compare these features with traditional website fingerprinting features in both DoH
traffic and web traffic over HTTPS.

\para{Evaluation in closed and open worlds.}
We first evaluate the attack in a closed world using the \loca dataset. We
try three settings: an adversary that attacks the full dataset of \cwpages websites, 
an adversary attacking the curated dataset of 1,414 websites after we eliminate spurious errors, 
and an adversary attacking the full dataset but that considers regional versions of
given pages to be equivalent. For example, classifying \texttt{google.es} as \texttt{google.co.uk},
a common error in our classifier, is considered a true positive.
We see in Table~\ref{tab:classification} that testing on the clean dataset offers just a 1\% 
performance increase, and that considering regional versions as equivalent results provides 
an additional 3\% increase. 
As in prior work on website fingerprinting~\cite{OverdorfJAGD17}, we have used confusion graphs.
Confusion graphs are an intuitive representation of classification errors that allows to visualize collisions between pages. 
See Figures~\ref{graph:outliers_wo_one_time_diff_len} and~\ref{graph:outliers_wo_one_time}
in the Appendix for the confusion graphs of this evaluation.
Given the minimal differences, in the reminder of the experiments we use the full dataset.

\begin{table}[t!]
  \centering
  \caption{Classifier performance for \loca dataset (mean and standard deviation for 10-fold cross validation).}
  \resizebox*{\columnwidth}{!}{%
  \begin{tabular}[c]{ r  c  c  c  c }
   \toprule
    \textbf{Scenario}  & \textbf{Precision} & \textbf{Recall} & \textbf{F1-score}\\
    \midrule
     Curated traces & $0.914 \pm 0.002$ & $0.909 \pm 0.002$ & $0.908 \pm 0.002$ \\
     Full dataset & $0.904 \pm 0.003$ & $0.899 \pm 0.003$ & $0.898 \pm 0.003$ \\
     Combined labels & $0.940 \pm 0.003$ & $0.935 \pm 0.003$ & $0.934 \pm 0.003$ \\
    \bottomrule
  \end{tabular}}
  \label{tab:classification}
\end{table}

\begin{figure}[t!]
\centering
\includegraphics[width=0.45\textwidth]{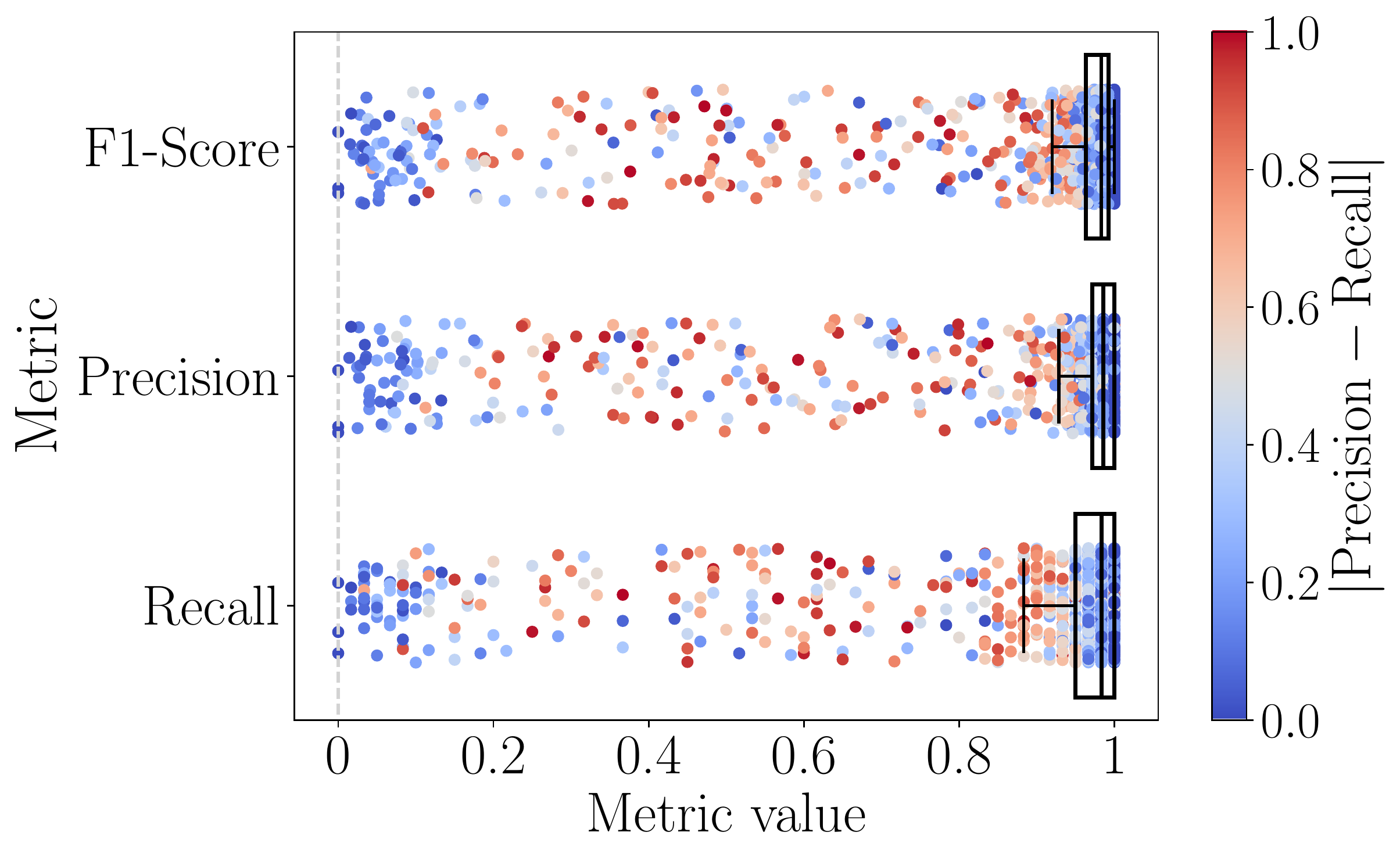}
\caption{Performance per class in \loca. Each dot represents 
a class and its color the absolute difference between Precision and Recall (blue low, red high).}
\label{fig:got_boxplots}
\vspace{-0.7cm}
\end{figure}

% GOT scatterplots
In the context of website fingerprinting, Overdorf \etal~\cite{OverdorfJAGD17}
showed that it is likely that the classifier's performance varies
significantly between different individual classes. Thus, looking
only at average metrics, as in Table~\ref{tab:classification}, may
give an incomplete and biased view of the classification results. 
To check if this variance holds on DNS traces we study the 
performance of the classifier for individual websites. The result is shown in Figure~\ref{fig:got_boxplots}.
In this scatterplot each dot is a website and its color represents the absolute
difference between Precision and Recall: blue indicates 0 difference and red indicates maximum
difference (i.e., $|Precision-Recall| = 1$). 
We see that some websites (red dots on the right of the Precision scatterplot) 
have high Recall -- they are often identified by the classifier,
but low Precision -- other websites are also identified as the this website. 
Thus, these websites have good privacy since the false positives provide the 
users with plausible deniability.
For other pages (red dots on the right of the Recall scatterplot),
the classifier obtains low Recall -- it almost never identifies them,
but high Precision -- if they are identified, the adversary is absolutely sure
her guess is correct. The latter case is very relevant for privacy, 
and in particular censorship, as
it enables the censor to block without fear of collateral damage.

\emph{Open world.}
In the previous experiments, the adversary knew that the webpage visited by the victim was 
within the training dataset. We now evaluate the adversary's
capability to distinguish those webpages from other unseen traffic.
Following prior work~\cite{JansenJGED18,
JuarezIPDW16} we consider two sets of webpages, one \textit{monitored} and 
one \textit{unmonitored}.
The adversary's goal is to determine whether a test trace belongs to a page within
the monitored set. 

We train a classifier with both monitored and unmonitored samples. 
Since it is not realistic to assume that an adversary can have access to 
all unmonitored classes, we create unmonitored samples using 5,000 webpages traces 
formed by a mix of the \ow and \loca datasets. We divide the classes
such that 1\% of all classes are
in the monitored set and 10\% of all classes are used for training. 
We ensure that the training dataset is balanced, i.e., 
it contains equal number of monitored and unmonitored samples; and the 
test set contains an equal number of samples from classes used in training and 
classes unseen by the classifier. 
When performing cross validation, in every fold 
we consider a different combination of the 
monitored and unmonitored classes for training and testing so that we
do not overfit to a particular case. 

\begin{figure}[t!]
\centering
\includegraphics[width=1\linewidth]{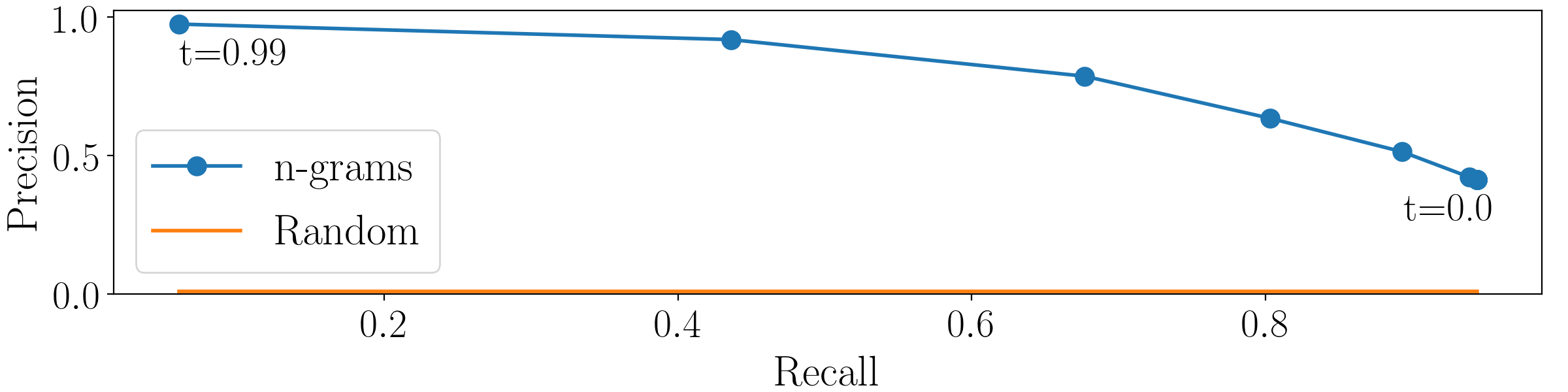}
\caption{Precision-Recall ROC curve for open world classification, for the monitored class. 
The threshold, $t$, 
is varied from 0.0 to 0.99 in steps of 0.1 (standard deviation less than 1\%).}
\label{fig:prp1}
	\vspace{-0.7cm}
\end{figure}

To decide whether a target trace is monitored or unmonitored,
we use a method proposed by Stolerman et al.~\cite{StolermanBreaking14}.
We assign the target trace to the monitored class if and only if the classifier predicts 
this class with probability larger than a threshold $t$, and to unmonitored otherwise.
We show in Figure~\ref{fig:prp1}, the average Precision-Recall ROC curve for the monitored 
class over 10 iterations varying the discrimination threshold, $t$, from 0 to 0.99 in steps of 0.1. 
We also show the random classifier, which indicates the probability of selecting the 
positive class uniformly at random, and acts as a baseline.
We see that when $t = 0.8$, the classifier has an F1-score of $\approx$ 0.7.
This result suggests that traffic analysis is a true threat to DNS privacy.

\para{Comparison to Web traffic fingerprinting.}
To understand the gain DNS fingerprinting provides to an adversary, we
compare its effectiveness to that of Web traffic fingerprinting. We also
evaluate the suitability of n-grams and traditional website fingerprinting features to
both fingerprinting problems.
We compare it to state-of-the-art attacks that use different features: the k-Fingerprinting attack, by Hayes and Danezis~\cite{HayesD16}, that considers a comprehensive set of features used in the website fingerprinting literature; CUMUL, by Panchenko \etal~\cite{PanchenkoLZHPWE16} which focuses on packets' lengths and order through cumulative features; and Deep Fingerprinting (DF), an attack based on deep convolutional neural networks~\cite{SirinamIJW18}.
In this comparison, we consider a closed-world of 700 websites (\wfp dataset) and use a random forest with the same parameters as classification algorithm.
We evaluate the performance of the classifiers on only DoH traffic (DoH-only), only HTTPS traffic corresponding to web content traffic (Web-only), and a mix of the two in order to verify
the claim in the RFC that DoH~\cite{doh} holds great potential to thwart traffic analysis. Table~\ref{tab:webfp_analysis} shows the results.

First, we find that for DNS traffic, due to its chatty characteristics, n-grams provide more than 10\% performance improvement with respect to traditional features. 
We also see that the most effective attacks are those made on web traffic. This is not surprising, as the variability of resources' sizes in web traffic contains more information than the small DNS packets. What is surprising is that the n-grams features \emph{outperform} the traditional features \emph{also} for website traffic. Finally, as predicted by the standard, if DoH and HTTPS are sent on the same TLS tunnel and cannot be separated, both set of features see a decrease in performance with n-grams offering ~10\% improvement. 

In summary, the best choice for an adversary with access to the isolated HTTPS flow is to analyse that trace with our novel n-grams features. However, if the adversary is in `a different path than the communication between the initiator and the recipient'~\cite{rfc7626} where she has access to DNS, or is limited in resources (see below), the DNS encrypted flow provides comparable results.

\begin{table}[t!]
  \centering
  \caption{F1-Score of the n-grams, k-Fingerprinting, CUMUL and DF features for different subsets of traffic: only DoH traffic (DoH-only), only HTTPS traffic corresponding to web traffic (Web-only) and mixed (DoH+Web).}
  \resizebox*{0.95\columnwidth}{!}{%
\begin{tabular}{cccc}
  \toprule
									& DoH-only		&  Web-only		& DoH + Web\\
  \midrule
n-grams								& 0.87			& 0.99			& 0.88 \\
k-Fingerprinting~\cite{HayesD16}    & 0.74			& 0.95			& 0.79 \\
CUMUL~\cite{PanchenkoLZHPWE16}		& 0.75			& 0.92			& 0.77 \\
DF~\cite{SirinamIJW18}\tablefootnote{We evaluate DF's accuracy following the original paper, i.e., using validation and test sets, instead of 10-fold cross-validation.}				& 0.51			& 0.94			& 0.75 \\
  \bottomrule
  \end{tabular}}
  \label{tab:webfp_analysis}
\end{table}

\para{Adversary's effort.} 
An important aspect to judge the severity of traffic analysis attacks is the effort needed 
regarding data collection to train the classifier~\cite{NasrHM17}. We study this effort from
two perspectives: amount of samples required -- which relates to the time needed to 
prepare the attack, and volume of data -- which relates to the storage and processing 
requiremets for the adversary.

We first look at how many samples are required to train a well-performing classifier.  
We see in Table~\ref{tab:samples} that there is a small increase between 10 and 20 samples, 
and that after 20 samples, there are diminishing returns in 
increasing the number of samples per domain. 
This indicates that, in terms of number of samples, the collection effort to
perform website identification on DNS traces is \emph{much smaller} 
than that of previous work on web traffic analysis: 
Website fingerprinting studies in Tor report more than 10\% increase between 10 and 20
samples~\cite{WangG13} and between 2\% and 10\% between 100 and 200
samples~\cite{RimmerPJVW18,SirinamIJW18}.

We believe the reason why fingerprinting DoH requires fewer samples per domain
is DoH's lower intra-class variance with respect to encrypted web traffic.
This is because sources of large variance in web traffic, such as the
presence of advertisements which change accross visits thereby varying the sizes of the resources,
does not show in DNS traffic for which the same ad-network domains are resolved~\cite{bashir2018diffusion}.

\begin{table}[t!]
  \centering
  \footnotesize
  \caption{Classifier performance for different number of samples in the 
  \loca dataset averaged over 10-fold cross validation (standard deviations less than 1\%).}
  \begin{tabular}[c]{ c  c  c  c }
   \toprule
   \textbf{Number of samples}  & \textbf{Precision} & \textbf{Recall} & \textbf{F1-score}\\
	 %  &           &            &         \\
	\midrule
	10         &     0.873 &      0.866 &   0.887 \\
	20         &     0.897 &      0.904 &   0.901 \\
	40         &     0.908 &      0.914 &   0.909 \\
	100        &     0.912 &      0.916 &   0.913 \\
	%200        &     0.905 &      0.908 &   0.906 \\
    \bottomrule
  \end{tabular}
  \label{tab:samples}
\end{table}

In terms of volume of data required to launch the attacks, targeting DoH flows also brings great advantage.
In the \wfp dataset, we observe that web traces have a length of 1.842 MB $\pm$ 2.620 MB,
while their corresponding DoH counterpart only require 0.012 MB $\pm$ 0.012 MB. While this
may not seem a significant difference, when we look at the whole dataset instead of individual traces,
the HTTPS traces require 73GB while the DoH-only dataset fits in less than 1GB (0.6GB).
This is because DNS responses are mostly small, while web traffic request and
responses might be very large and diverse (\eg different type of resources, or
different encodings).

In our experiments, to balance data collection effort and performance, 
we collected 60 samples per domain for all our datasets. For the unmonitored websites 
in the open world, for which we do not need good models as we do not
identify them, we collected three samples per domain.

\subsection{DNS Fingerprinting Robustness}
\label{sec:robustness}

In practice, the capability of the adversary to distinguish
websites is very dependent on differences between the environmental 
conditions at attack time and the setup for training data collection~\cite{JuarezAADG14}. 
We present experiments exploring three environmental dimensions: time, space, and infrastructure.

\subsubsection{Robustness over time}
\label{sec:time}

DNS traces vary due to the dynamism of webpage content and variations in DNS responses (e.g.,
service IP changes because of load-balancing). To understand the impact of this variation
on the classifier, we consider collect data \loca for 10 weeks from the end 
of September to the beginning of November 2018. 
We divide this period into five intervals, each containing two 
consecutive weeks, and report in Table~\ref{tab:time} the F1-score of the classifier when we 
train the classifier on data from a single interval and use the other intervals as test data 
(0 weeks old denotes data collected in November). 
In most cases, the F1-score does not significantly decrease within a period of
4 weeks. Longer periods result in a significant drops -- more than 10\% drop in F1-score
when the training and testing are separated by 8 weeks.

\begin{table}[t!]
  \centering
	\caption{F1-score when training
	on the interval indicated by the row and testing on the interval in the
	column (standard deviations less than 1\%). We use 20 samples per webpage (the maximum number of samples collected
in all intervals).}
  \resizebox*{\columnwidth}{!}{%
	\begin{tabular}[c]{ c  c  c  c  c  c}
	\toprule
			\textbf{F1-score}	& \textbf{0 weeks old}  & \textbf{2 weeks old}  & \textbf{4 weeks old} & \textbf{6 weeks old} & \textbf{8 weeks old}\\
		\midrule
			\textbf{0 weeks old} & \cTab{0.880} & \cTab{0.827} & \cTab{0.816} & \cTab{0.795} & \cTab{0.745} \\
			\textbf{2 weeks old} & \cTab{0.886} & \cTab{0.921} & \cTab{0.903} & \cTab{0.869} & \cTab{0.805} \\
			\textbf{4 weeks old} & \cTab{0.868} & \cTab{0.898} & \cTab{0.910} & \cTab{0.882} & \cTab{0.817} \\
			\textbf{6 weeks old} & \cTab{0.775} & \cTab{0.796} & \cTab{0.815} & \cTab{0.876} & \cTab{0.844} \\
			\textbf{8 weeks old} & \cTab{0.770} & \cTab{0.784} & \cTab{0.801} & \cTab{0.893} & \cTab{0.906} \\
		\bottomrule
	\end{tabular}}
  \label{tab:time}
\end{table}

This indicates that, to obtain best performance, an adversary with the resources of a university research group would need to collect data at least once a month.
However, it is unlikely that DNS traces change drastically. To account for gradual changes, 
the adversary can perform continuous collection and mix data across weeks. 
In our dataset, if we combine two- and three-week-old samples for training; 
we observe a very small decrease in performance. Thus, a
continuous collection strategy can suffice to maintain the adversary's performance without
requiring periodic heavy collection efforts.

\subsubsection{Robustness across locations}
\label{sec:loc}
DNS traces may vary across locations due to several reasons.
First, DNS lookups vary when websites adapt 
their content to specific geographic regions. Second, 
popular resources cached by resolvers vary across regions.
Finally, resolvers and CDNs 
use geo-location methods for load-balancing requests,
\eg using anycast and EDNS~\cite{otto2012content, rula2014behind}. 
 
We collect data in three locations, two countries in Europe (\loca and \locb) 
and a third in Asia (\locc). 
Table~\ref{tab:changes} (leftmost) shows the classifier performance when 
crossing these datasets for training and testing.
When trained and tested on the same location unsurprisingly the classifier 
yields results similar to  the ones obtained in the base experiment. When we 
train and test on different locations, the F1-score decreases between a 16\% and a
27\%, the greatest drop happening for the farthest location, \locc, in Asia.

Interestingly, even though \locb yield similar F1-Scores when
cross-classified with \loca and \locc, the similarity does not hold when looking 
at Precision and Recall individually.
For example, training on \locb and testing on \loca results on around 77\% Precision and Recall,
but training on \loca and testing on \locb yields 84\% Precision and 65\% Recall. 
Aiming at understanding the reasons behind this asymmetry, we build a classifier
trained to separate websites that obtain high recall (top 25\% quartile) and low 
recall (bottom 25\% quartile) when training with \loca and \locc and testing in \locb. 
A feature importance analysis on this classifier that \locb's low-recall top features 
have a significantly lower importance in \loca and \locb. Furthermore, 
we observe that the intersection between \loca and \locc's relevant feature sets is 
slightly larger than their respective intersections with \locb. While it is 
clear that the asymmetry is caused by the configuration of the network in \locb, 
its exact cause remains an open question.

\subsubsection{Robustness across infrastructure}
\label{sec:infra}

In this section, we study how the DoH resolver and client,
and the user platform affect influence the attack's performance.

\para{Influence of DoH Resolver.}
\label{sec:resolver}
We study two commercial DoH resolvers, Cloudflare's and Google's.
Contrary to Cloudflare, Google does not provide a stand-alone DoH client.
To keep the comparison fair, we instrument a new collection setting using Firefox in its 
\emph{trusted recursive resolver} configuration with both
DoH resolvers. 

\begin{table*}[t!]
  \centering
  \caption{Performance variation changes in location and infrastructure (F1-score, standard deviations less than 2\%).} 
  	\resizebox*{0.48\columnwidth}{!}{%
  	\begin{tabular}[t]{ c c c c}
  		\toprule
			\textbf{Location}	& \textbf{\loca} 	& \textbf{\locb} 	& \textbf{\locc} \\
		\midrule
		  \textbf{\loca}		& \cTab{0.906}		& \cTab{0.712}		& \cTab{0.663} \\
		  \textbf{\locb}		& \cTab{0.748}		& \cTab{0.908}		& \cTab{0.646} \\	
		  \textbf{\locc}		& \cTab{0.680}		& \cTab{0.626}		& \cTab{0.917} \\
		\bottomrule
	\end{tabular}}\hfill
  	\resizebox*{0.48\columnwidth}{!}{%
	\begin{tabular}[t]{ c c c}
	   \toprule
			\textbf{Resolver}	& \textbf{\google} 		& \textbf{\cloud} \\
	   \midrule
		  \textbf{\google}		& \cTab{0.880}			& \cTab{0.129} \\
		  \textbf{\cloud}		& \cTab{0.862}			& \cTab{0.885} \\
		\bottomrule
	\end{tabular}}\hfill
	\resizebox*{0.48\columnwidth}{!}{%
	\begin{tabular}[t]{ c c c}
	\toprule
			\textbf{Platform}		& \textbf{\desktop} 		& \textbf{\rpi} \\
		\midrule
		  \textbf{\desktop}		& \cTab{0.8802}			& \cTab{0.0003} \\
		  \textbf{\rpi}		& \cTab{0.0002}			& \cTab{0.8940} \\
		\bottomrule
	\end{tabular}}\hfill
	\resizebox*{0.48\columnwidth}{!}{%
	\begin{tabular}[t]{ c c c c}
		\toprule
			\textbf{Client}	& \textbf{\clia} 	& \textbf{\clib} 	& \textbf{\clic} \\
		\midrule
		  \textbf{\clia}		& \cTab{0.885}		& \cTab{0.349}		& \cTab{0.000} \\
		  \textbf{\clib}		& \cTab{0.109}		& \cTab{0.892}		& \cTab{0.069} \\	
		  \textbf{\clic}		& \cTab{0.001}		& \cTab{0.062}		& \cTab{0.908} \\
		\bottomrule
	\end{tabular}}
  \label{tab:changes}
\end{table*}

Table~\ref{tab:changes} (center-left) shows the result of the comparison.
As expected, training and testing on the same resolver yields the best results.
In particular, we note that even though Google hosts other services behind its resolver's IP
and thus DoH traffic may be mixed with the visited website's traffic (e.g., if a 
web embeds Google third-party) the classifier performs equally for both resolvers.

As in the location setting, we observe an asymmetric decrease 
in one of the directions: training on \google dataset
and attacking \cloud results in 13\% F1-score, while attacking \google 
with a classifier trained on \cloud yields similar results as training on \google itself.

\begin{figure}[t!]
\centering
\includegraphics[width=1\linewidth]{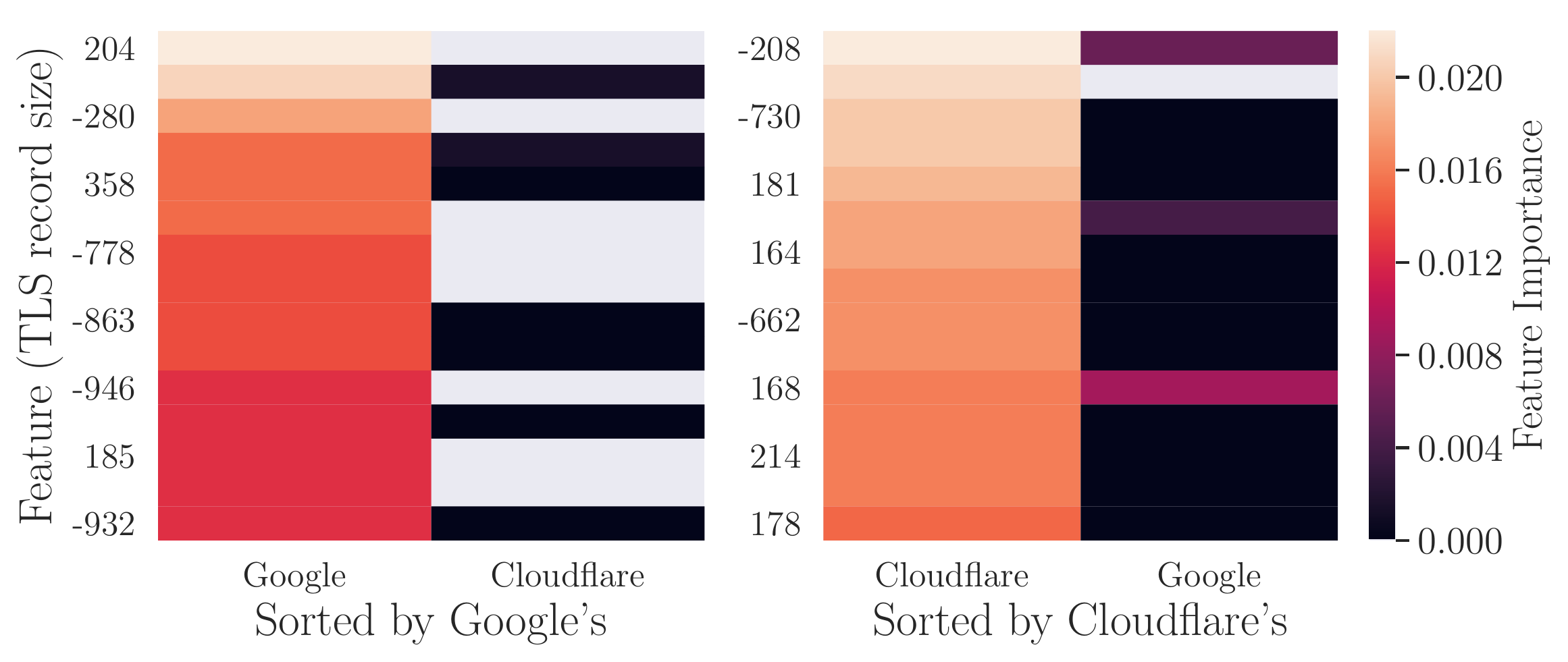}
\caption{Top 15 most important features in Google's and
	Cloudflare's datasets. On the left, features are sorted by the results
	on Google's dataset and, on the right, by Cloudflare's.}
\label{fig:heatmap_resolver}
	\vspace{-0.7cm}
\end{figure}

To investigate this asymmetry we rank the features according their
importance for the classifiers.
For simplicity, we only report the result on length unigrams,
but we verified that our conclusions hold when considering
all features together.
Figure~\ref{fig:heatmap_resolver} shows
the top-15 most important features for a classifier trained on Google's resolver (left) and
Cloudflare's (right). 
The rightmost diagram of each column shows the importance of these features
on the other classifier. Red tones indicate high importance, and dark colors 
represent irrelevant features. Grey indicates that the feature is not present.

We see that the most important features in Google are either not important or missing in Cloudflare 
(the right column in left-side heatmap is almost gray). 
% Recall that our features are
% the counts of TLS record sizes. A missing feature translates into zero count. 
As the missing features are very important, they induce erroneous splits
early in the trees, and for a larger fraction of the data, causing the performance drop.
However, only one top feature in the classifier trained on Cloudflare
is missing in Google, and the others are also important
(right column in right-side heatmap). Google does miss important features 
in Cloudflare, but they are of little importance and
their effect on performance is negligible.

\para{Influence of end user's platform.}
%Next we study influence of the platform in which the DoH client runs.
We collect traces for the 700 top Alexa webpages on a Raspberry Pi (\rpi dataset) and 
an Ubuntu desktop (\desktop dataset), both from \loca.
% We use traces collected on an Ubuntu desktop (the \loca dataset) and on a Raspberry Pi (the \rpi
% dataset). Since \rpi only has 700 webpages, we select a subset of the same 700 webpages from \loca
% which we denote as \desktop. 
We see in Table~\ref{tab:changes} (center-right) that, as expected, the classifier has good performance 
when the training and testing data come from the same platform. 
However, it drops to almost zero when crossing the datasets.
% Thinking that these could be small classification errors we looked at the top-k candidates to determine if the correct
% webpages were among the most likely predictions. However, only 13\% of the correct choices are among the top-100 candidates. 

When taking a closer look at the TLS record sizes from both platforms we found that TLS records in the \desktop dataset are on average 7.8 bytes longer than those in
\rpi (see Figure~\ref{fig:histplatform} in Appendix~\ref{sec:extra}). We repeated the cross classification after adding 8 bytes to all \rpi TLS record sizes. 
Even though the classifiers do not reach the base experiment's performance, we see a significant improvement in
cross-classification F1-score to $0.614$ when training on \desktop and testing on \rpi, and $0.535$ 
when training on \rpi and testing on \desktop. 
% \begin{table}[htbp]
%   \centering
%   \caption{Performance when training on the platform indicated by the row and testing on the platform
%   indicated in the column (standard deviation less than 1\% for same platform and less than 0.1\% 
%   for cross-platform.}
%   \resizebox*{0.48\columnwidth}{!}{%
% 	\begin{tabular}[c]{ c  c  c}
% 	\toprule
% 			\textbf{F1-Score	}		& \textbf{\desktop} 		& \textbf{\rpi} \\
% 		\midrule
% 		  \textbf{\desktop}		& \cTab{0.8802}			& \cTab{0.0003} \\
% 		  \textbf{\rpi}		& \cTab{0.0002}			& \cTab{0.8940} \\
% 		\bottomrule
% 	\end{tabular}}
%   \label{tab:platform}
% \end{table}

\para{Influence of DNS client.}
Finally, we consider different client setups: Firefox's trusted recursive resolver or TRR (\clia),
Cloudlflare's DoH client with Firefox (\clib) and Cloudflare's DoH client with Chrome (\clic).
We collected these datasets in location \locb using Cloudflare's resolver.
 %differences in classifier performance
%when we use different clients for training and testing (see Table~\ref{tab:client}).

% \begin{table}[htbp]
%   \centering
% \caption{Performance when training on the client setups indicated by the row 
% and testing on the configuration indicated by the column
% (standard deviations less than 2\%).}
%   \resizebox*{0.49\columnwidth}{!}{%
% 	\begin{tabular}[c]{ c c c c}
% 		\toprule
% 			\textbf{F1-Score}	& \textbf{\clia} 	& \textbf{\clib} 	& \textbf{\clic} \\
% 		\midrule
% 		  \textbf{\clia}		& \cTab{0.885}		& \cTab{0.349}		& \cTab{0.000} \\
% 		  \textbf{\clib}		& \cTab{0.109}		& \cTab{0.892}		& \cTab{0.069} \\	
% 		  \textbf{\clic}		& \cTab{0.001}		& \cTab{0.062}		& \cTab{0.908} \\
% 		\bottomrule
% 	\end{tabular}}
%   \label{tab:client}
% \end{table}

Table~\ref{tab:changes} (rightmost) shows that the classifier performs as expected
when trained and tested on the same client setup.
When the setup changes, the performance of the
classifier drops dramatically, reaching zero when we use different
browsers. We hypothesize that the decrease between
\clib and \clic is due to differences in the implementation of the
Firefox's built-in and Cloudflare's standalone DoH clients.

Regarding the difference when changing browser, we found that Firefox' 
traces are on average 4 times longer than Chrome's. 
We looked into the unencrypted traffic to understand
this difference. 
We used a proxy to man-in-the-middle the DoH connection between the client and the
resolver\footnote{https://github.com/facebookexperimental/doh-proxy}, obtaining
the OpenSSL TLS session keys with Lekensteyn's scripts\footnote{https://git.lekensteyn.nl/peter/wireshark-notes}. 
We use this proxy to decrypt DoH captures for Firefox configured to use Cloudflare's
resolver, but we could not do the same for Google. Instead, we 
man-in-the-middle a curl-doh client\footnote{https://github.com/curl/doh},
which also has traces substantially shorter than Firefox. We find that Firefox, besides resolving domains 
related to the URL we visit, also issues resolutions related to OSCP servers, captive portal
detection, user's profile/account, web extensions, and other Mozilla servers.
As a consequence, traces in \ff and \cloud datasets are substantially larger 
and contain contain different TLS record sizes than any of 
our other datasets. We conjecture that Chrome performs similar requests, but since
traces are shorter we believe the amount of checks seems to be smaller than Firefox's. 
%  This results in more
% background noise for Firefox and substantially larger traces, which seems to
% impact the classifier.

\subsubsection{Robustness Analysis Takeaways}
Our robustness study shows that to obtain best results
across different configurations the adversary needs i) to train
a classifier for each targeted setting, and ii) to be 
able to identify her victim's configuration. 
Kotzias et al.\ demonstrated that identifying client or resolver is possible, 
for instance examining the IP (if the IP is dedicated to the resolver), 
or fields in the ClientHello of the TLS connection (such as the 
the Server Name Indication (SNI), cipher suites ordering, etc.)~\cite{KotziasRAPVC18}.
Even if in the future these features are not available, we found that the characteristics of 
the traffic itself are enough to identify a resolver. We built classifiers to distinguish 
resolver and client based on the TLS record length. We can identify
resolvers with 95\% accuracy, and we get no errors (100\% accuracy) when identifying the client.

Regarding users' platform, we see little difference
between desktops, laptops, and servers in Amazon Web Services. When the devices are as different
as a desktop and a constrained device such a Raspberry Pi, the classifier's accuracy drops.

Finally, our longitudinal analysis reveals that, keeping up with the changes in DNS traces can be done at 
low cost by continuously collecting samples and incorporating them to the training set.

\begin{figure}[t!]
\centering
\includegraphics[width=0.4\textwidth]{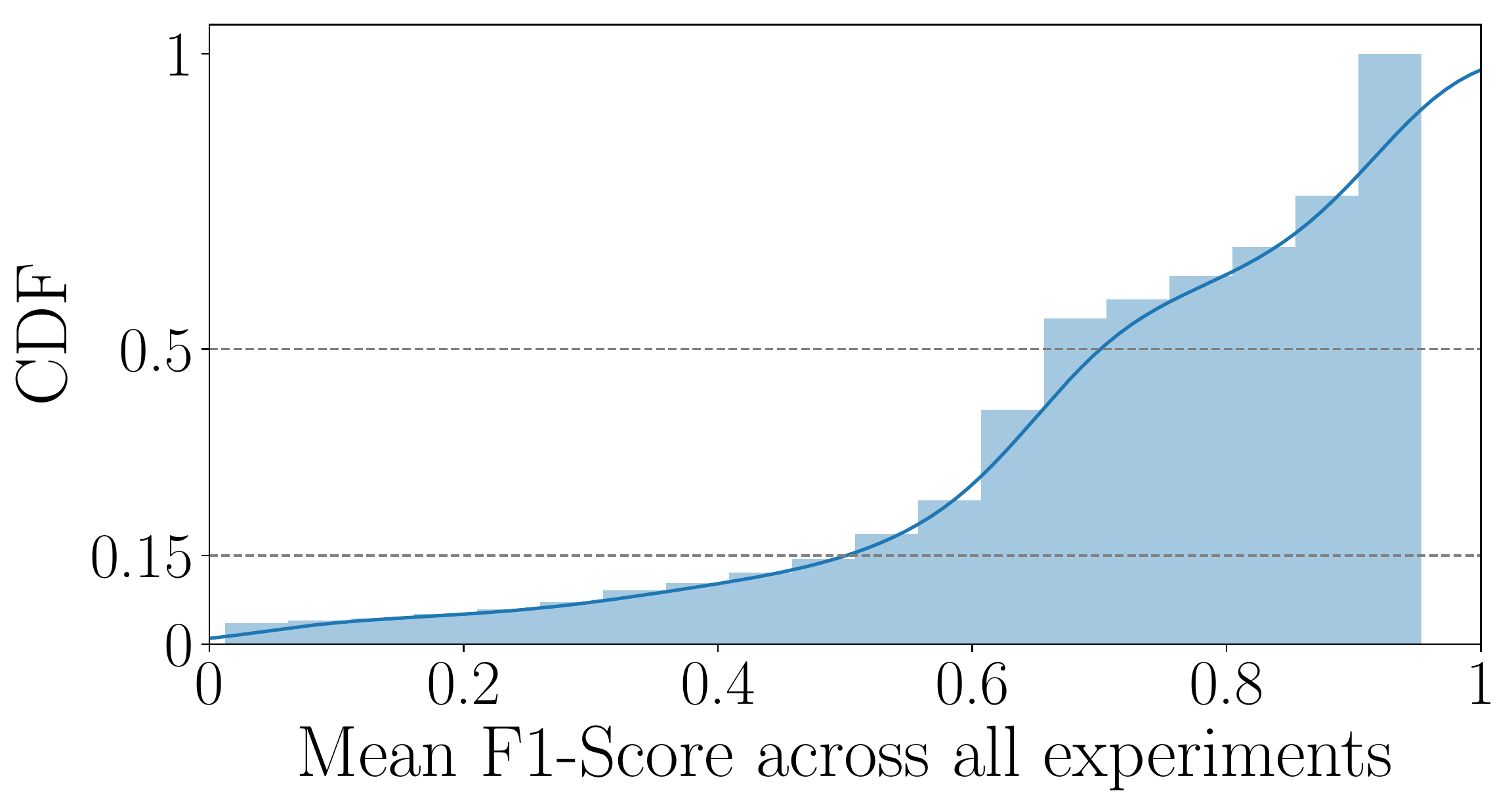}
    \caption{Cumulative Distribution Function (CDF) of the per-class mean F1-Score. }
\label{fig:cdf_survivors}
	\vspace{-0.7cm}
\end{figure}

\para{Survivors and Easy Preys.}
Finally, we study whether there are websites that are particularly good or bad at
evading fingerprinting under all the configurations. We compute the mean F1-Score across all configurations as
an aggregate measure of the attack's overall performance.
We plot the CDF of the distribution of mean F1-scores over the websites in
Figure~\ref{fig:cdf_survivors}. This distribution is heavily skewed: there are
up to 15\% of websites that had an F1-Score equal or lower than 0.5 and more
than 50\% of the websites have a mean F1-Score equal or lower than 0.7.

We then rank sites by lowest mean F1-Score and lowest standard deviation. 
At the top of this ranking, there are sites that \emph{survived} the attack in all configurations. 
Among these survivors, we found Google, as well as sites giving errors, that get 
consistently misclassified as each other. We could not find 
any pattern in the website structure or the resource loads that explains why other
sites with low F1 survive. We leave a more in-depth analysis of the survivors for future
work. At the bottom of the ranking, we find sites with long domain names, with few resources and that
upon visual inspection seem to present low variability.
In Tables~\ref{tab:doomed},~\ref{tab:survivors},~\ref{tab:high_variance} in the Appendix, we list the top-10 sites in the tails of
the distribution.

\section{DNS Defenses against fingerprinting}
\label{sec:countermeasures}
%!TEX root = report.tex
\looseness=-1
In this section, we evaluate the effectiveness of existing techniques to
protect encrypted DNS against traffic analysis. Table~\ref{tab:countermeasures}
summarizes the results. We consider the following defenses:

\parait{EDNS(0) Padding.} 
EDNS (Extension mechanisms for DNS) is a specification to increase the functionality of 
the DNS protocol~\cite{edns0}. It specifies how to add \emph{padding}~\cite{edns0padding}, 
both on DNS clients and resolvers, to prevent size-correlation attacks on encrypted DNS. 
The recommended padding policy is for clients to pad DNS requests to the nearest multiple of 128 bytes, 
and for resolvers to pad DNS responses to the nearest multiple of 468 bytes~\cite{paddingpolicies}.

Cloudflare's DoH client provides functionality to set EDNS(0) padding 
to DNS queries, leaving the specifics of the padding policy to the user. 
We modify the client source code to follow the padding strategy above. Google's specification
also mentions EDNS padding. However, we could not find any option to activate this feature, thus
we cannot analyze it.

In addition to the EDNS(0) padding, we wanted to see whether simple user-side measures that
alter the pattern of requests, such as the use of an ad-blocker, could be effective countermeasures.
We conduct an experiment where we use Selenium with a Chrome instance with the AdBlock Plus 
extension installed. We do this with DoH requests and responses padded to multiples of 128 bytes.

Upon responsible disclosure of this work, Cloudflare's added padding to
the responses of their DoH resolver. However, when analyzing the collected data we discover that
they do \emph{not} follow the recommended policy. Instead of padding to multiples of 468 bytes,
Cloudflare's resolver pads responses to multiples of 128 bytes, as recommended for DoH clients.
In order to also evaluate the recommended policy, we set up an HTTPS proxy (\emph{mitmproxy}) between
the DoH client and the Cloudflare resolver. The proxy intercepts responses from Cloudflare's DoH resolver, 
strips the existing padding, and pads responses to the nearest multiple of 468 bytes. 

As we show below, none of these padding strategies completely stops traffic analysis.
To understand the limits of protection of padding, we simulate a setting in which 
padding is perfect, \ie \emph{all} records have the same length and the classifier 
cannot exploit the TLS record size information. To simulate this setting, we
artificially set the length of all packets in the dataset to 825, the maximum size
observed in the dataset.

\parait{DNS over Tor.}
Tor is an anonymous communication network. To protect the privacy of its users, Cloudflare
set up a DNS resolver that can be accessed using Tor. This enables users to not reveal their IP to 
the resolver when doing lookups. 

To protect users' privacy, Tor re-routes packets through so-called onion routers to 
avoid communication tracing based on IP addresses; and it packages content into 
constant-size cells to prevent size-based analysis. These countermeasures have so far not
been effective to protect web traffic~\cite{PanchenkoLZHPWE16, WangG16, HayesD16, SirinamIJW18}.
We study whether they can protect DNS traffic.

\para{Results.}
Our first observation is that EDNS0 padding is not as effective as expected. 
Adding more padding, as recommended in the specification, does provide better protection,
but still yields an F1-score of 0.45, six orders of magnitude greater than random guessing.
Interestingly, removing ads helps as much as increasing the padding, as shown by the
 EDNS0-128-adblock experiment.
As shown below, Perfect padding would actually deter the attack, but at a high communication
cost.

As opposed to web traffic, where website fingerprinting
obtains remarkable performance~\cite{PanchenkoNZE11, HayesD16, SirinamIJW18},
Tor is very effective in hiding the websites originating a DNS trace.
The reason is that DNS lookups and responses are fairly small. 
They fit in one, at most two, Tor cells which in turn materialize in few observed TLS record sizes.
As a result, it is hard to find features unique to a page. Also, 
DNS traces are shorter than normal web traffic, and present less variance. Thus, length-related features,
which have been proven to be very important in website fingerprinting, 
only provide a weak 1\% performance improvement.

Even though Perfect padding and DNS over Tor offer similar performance,
when we look closely at the misclassified webpages, we see that their 
behavior is quite different.
For Tor, we observe misclassifications cluster around six different groups, and in Perfect padding
they cluster around 12 (different) groups (Figures~\ref{graph:tor_misclassif_multiple} 
and~\ref{graph:const_padding_star} in the Appendix).
For both cases, we tested that it is possible to build a classifier that identifies
the cluster a website belongs to with reasonable accuracy. This means that
despite the large gain in protection with respect to EDNS(0), the effective anonymity 
set for a webpage is much smaller than the total number of webpages in the dataset.

\begin{table}[t!]
  \centering
  \footnotesize
  \caption{Classification results for countermeasures.}
   \resizebox*{\columnwidth}{!}{%
  \begin{tabular}[c]{ c c c c }
  \toprule
     \textbf{Method} & \textbf{Precision} & \textbf{Recall} & \textbf{F1-score}\\
    \midrule
    EDNS0-128 & $0.710 \pm 0.005$ & $0.700 \pm 0.004$ & $0.691 \pm 0.004$ \\
    EDNS0-128-adblock & $0.341 \pm 0.013$ & $0.352 \pm 0.011$ & $0.325 \pm 0.011$ \\
    EDNS0-468 & $0.452 \pm 0.007$ & $0.448 \pm 0.006$ & $0.430 \pm 0.007$ \\
    Perfect padding & $0.070 \pm 0.003$ & $0.080 \pm 0.002$ & $0.066 \pm 0.002$ \\
    DNS over Tor & $0.035 \pm 0.004$ & $0.037 \pm 0.003$ & $0.033 \pm 0.003$ \\
    \hline
    DNS over TLS & $0.419 \pm 0.008$ & $0.421 \pm 0.007$ & $0.395 \pm 0.007$ \\
    \bottomrule
  \end{tabular}}
  \label{tab:countermeasures}
\end{table}

\begin{figure}[t!]
   \footnotesize
   \centering
     \includegraphics[width=1\linewidth]{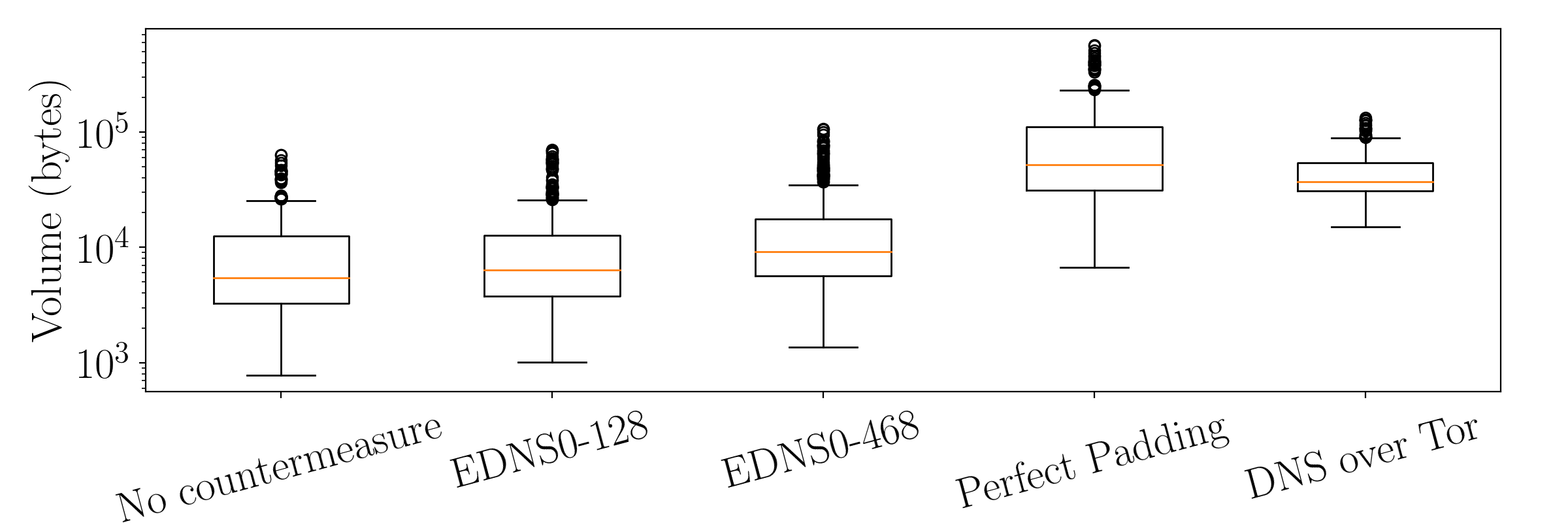}
   \caption{Total volume of traffic with and without countermeasures.}
   \label{fig:overhead}
	 \vspace{-0.7cm}
 \end{figure}

Finally, we evaluate defenses' communication overhead. 
For each countermeasure, we collect 10 samples of 50 webpages, with and without countermeasures,
and measure the difference in total volume of data exchanged between client and resolver.
We see in Figure~\ref{fig:overhead} that, as expected, EDNS0 padding (both 128 and 468) 
incur the least overhead, but they also offer the least protection. DNS over Tor,
in turn, offers lower overhead than Perfect padding.

 \para{Comparison with DNS over TLS (DoT).} Finally, we compare the protection
 provided by DNS over HTTPS and over TLS, using the \dot dataset. 
 As in DoH, Cloudflare's DoT resolver implements EDNS0 padding of DNS responses to a multiple of 128 bytes. 
 The cloudflared DoT client, however, does not support DoT traffic. Thus, we use the Stubby client to 
 query Cloudflare's DoT padded to a multiple of 128 bytes. 

\begin{figure}[h]
   \centering
     \includegraphics[width=1\linewidth]{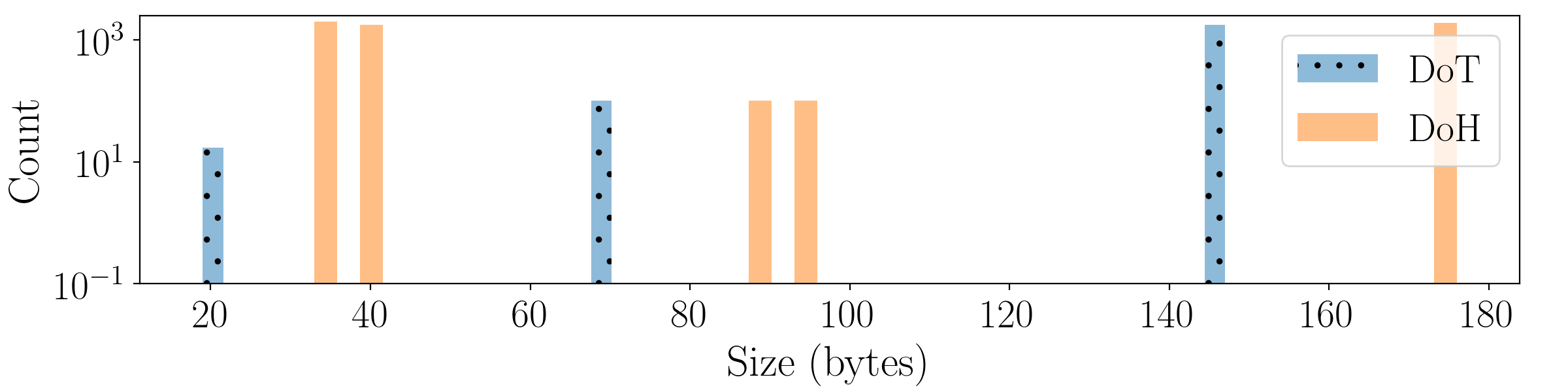}
   \\
     \includegraphics[width=1\linewidth]{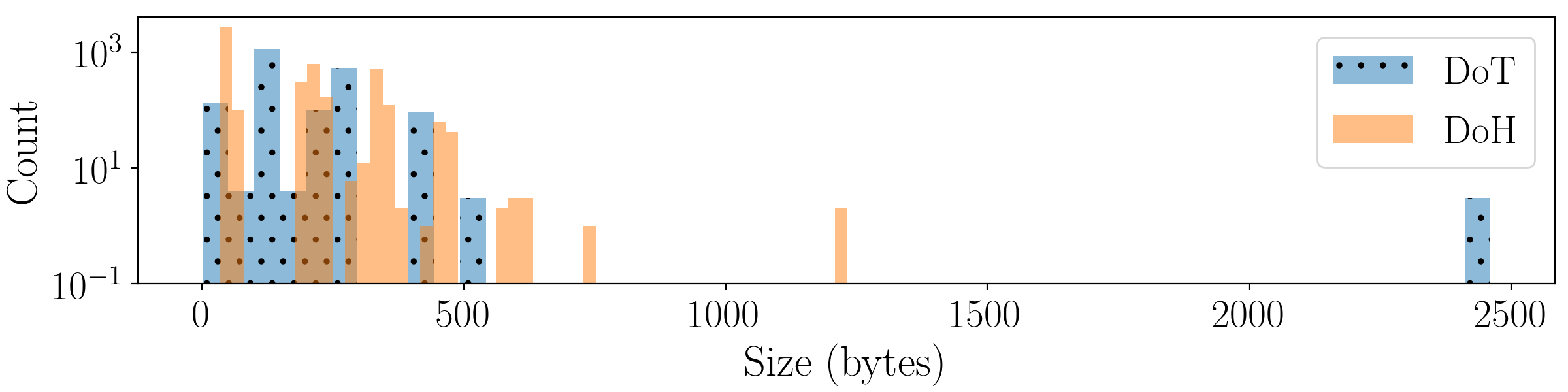}
   \caption{Histogram of sent (top) and received (bottom) TLS record sizes for DoT and DoH.}
   \label{fig:dot_doh}
	 \vspace{-0.7cm}
 \end{figure}

 Our results (shown in the last row of Table~\ref{tab:countermeasures}) indicate that DoT offers
 much better protection than DoH -- about 0.3 reduction in F1-score. The reason is that
 DoT traffic presents much less variance than DoH, and thus are less unique. We show this in Figure~\ref{fig:dot_doh},
 where we represent a histogram of sizes of the sent and received TLS records for both DoH and DoT for 100
 webpages. As both flows are encrypted, we cannot investigate the underlying reasons for this difference.
 
\section{Censorship on encrypted DNS}
\label{sec:censorship}
%!TEX root = report.tex

DNS-based blocking is a wide-spread method of censoring access to web
content. Censors inspect DNS lookups and, when they detect a blacklisted domain, 
they either reset the connection or inject their own DNS response~\cite{TschantzAP16}. 
The use of encryption in DoH precludes content-based lookup identification. 
The only option left to the censor is to block the resolver's IP. 
While this would be very effective, it is a very aggressive strategy, as 
it would prevent users from browsing any web. Furthermore, some DoH
resolvers, such as Google's, do not necessarily have a dedicated IP. Thus,
blocking their IP may also affect other services.

An alternative for the censor is to use traffic analysis to
identify the domain being looked up.
In this section, we study whether such an approach is 
feasible. We note that to block access to a site, a censor not only
needs to identify the lookup from the encrypted traffic, but also
needs to do this as soon as possible to prevent the user from downloading
any content. 

\subsection{Uniqueness of DoH traces}
In order for the censor to be able to uniquely identify domains given
DoH traffic, the DoH traces need to be unique. In particular, to enable
early blocking, \emph{the first packets} of the trace need to be unique. 

To study the uniqueness of DoH traffic, let us model the set of webpages in the world as a random variable $W$ with sample space $\Omega_W$; 
and the set of possible network traces generated by those websites as a random variable $S$ 
with sample space $\Omega_S$. A website's trace $w$ is a sequence of non-zero integers:
$(s_i)_{i=1}^n, s_i\in \mathbb{Z}\smallsetminus\{0\}$, $n\in \mathbb{N}$, where
$s_i$ represents the size (in bytes) of the $i$-th TLS record in the traffic
trace. Recall that its sign represents the direction -- negative for incoming (DNS to client) 
and positive otherwise. We denote partial traces, \ie only the $l$ first TLS 
records, as $S_l$.

We measure \emph{uniqueness} of partial traces using the conditional entropy $H(W \mid S_l)$, defined as:

$$H(W\mid S_l) = \displaystyle\sum_{\forall o\in \Omega_{S_l}} \Pr[S_l=o] H(W\mid
S_l=o)\,.$$
Here, $H(W\mid S_l=o)$ is the Shannon entropy of the probability
distribution $\Pr[W\mid S_l=o]$. This probability describes the likelihood that the adversary guesses 
websites in $W$ given the observation $o$. The conditional entropy $H(W\mid S_l)$
measures how distinguishable websites in $\Omega_W$ are when the adversary has only
observed $l$ TLS records. When this entropy is zero,
sites are perfectly distinct. For instance, if the first packet 
of every DoH trace had a different size, then the entropy $H(W\mid S_1)$ would be 0.

\begin{figure}[t!]
\centering
\includegraphics[width=0.48\textwidth]{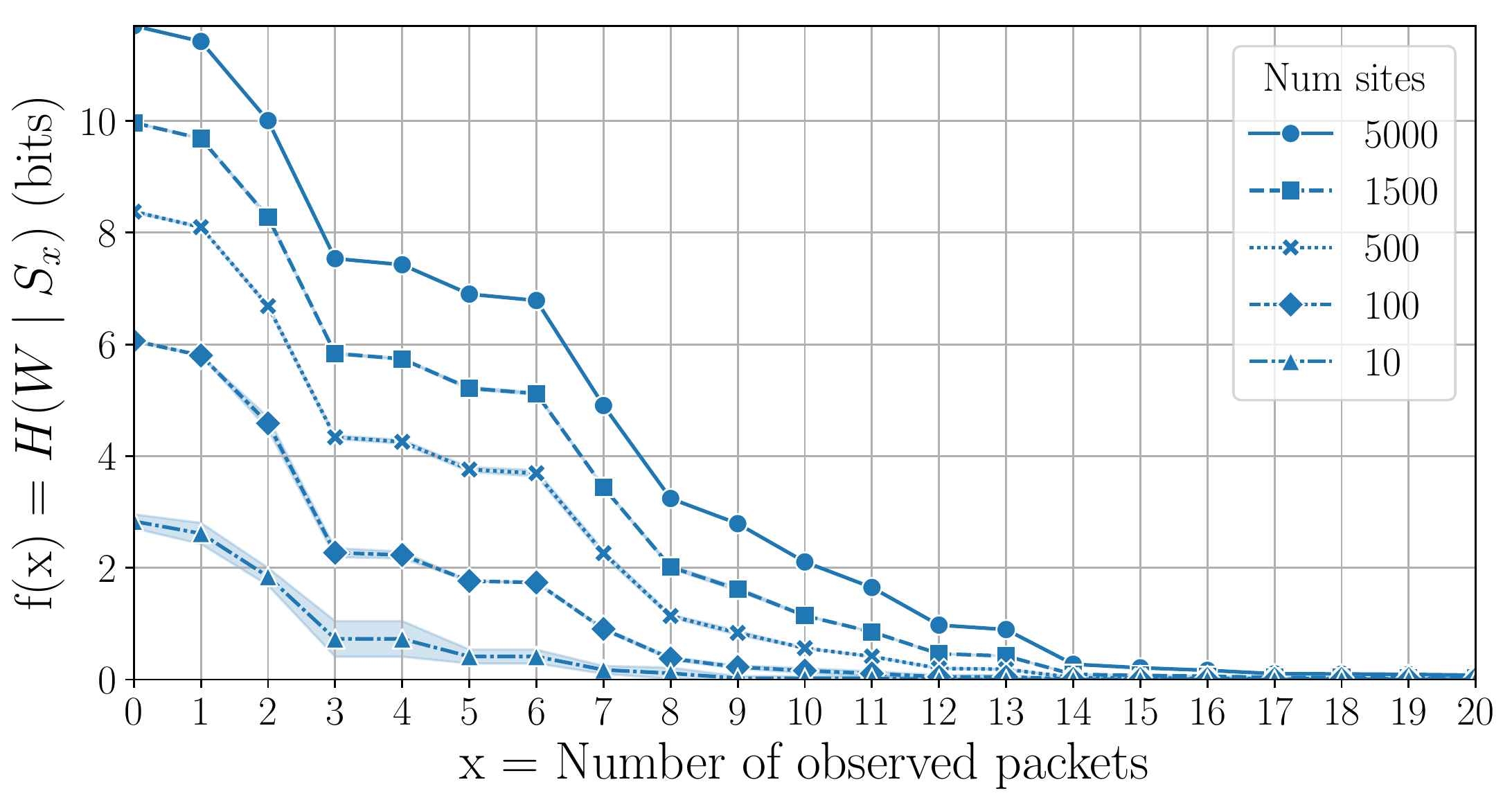}
\caption{Conditional entropy $H(W \mid S_l)$ given partial observations of DoH
	traces for different world sizes ($|\Omega_W|=\{10,100,500,1500,5000\}$). Each data point is averaged
	over 3 samples.}
\label{fig:entropies}
	 \vspace{-0.7cm}
\end{figure}

We compute the conditional entropy for different world sizes $|\Omega_W|$ and partial lengths $l$,
using traces from the \ow dataset. We show the result in Figure~\ref{fig:entropies}.
Every point is an average over 3 samples of $|\Omega_W|$ webs. These webs are selected uniformly 
at random from the dataset. The shades represent the standard deviation across the
3 samples.

Unsurprisingly, as the adversary observes more packets, the traces become more distinguishable
and the entropy decreases. For all cases, we observe a drop of up to 4 bits within the first
four packets, and a drop below 0.1 bits after 20 packets.
As the world size increases, the likelihood of having two or more 
websites with identical traces increases, and thus we observe a slower decay in entropy.
%% CARMELA: I reduced the discussion on variance since it does not seem really relevant.
We also observe that, as the world size increases the standard deviation becomes negligible.
This is because the \ow dataset contains 5,000 websites. Thus, as the size of the world increases, 
the samples of $\Omega_W$ contain more common websites. 

Even when considering 5,000 pages, the conditional entropy drops below 1 bit
after $15$ packets. This means that after $15$ packets have been observed, 
there is one domain whose probability of having generated the trace is larger
than 0.5. In our dataset $15$ packets is, on average, just 15\% of the whole trace. 
Therefore, on average,  
the adversary only needs to observe the initial 15\% of
a DoH connection to determine a domain with more confidence than taking a random guess
between two domains. We observe that as the number of pages increase, the curves 
are closer to each other indicating that convergence on uniqueness after the $15$-th packet 
is likely to hold in larger datasets.

\begin{figure}[t!]
\centering
\begin{subfigure}{0.49\textwidth}
  \centering
  \includegraphics[width=1\linewidth]{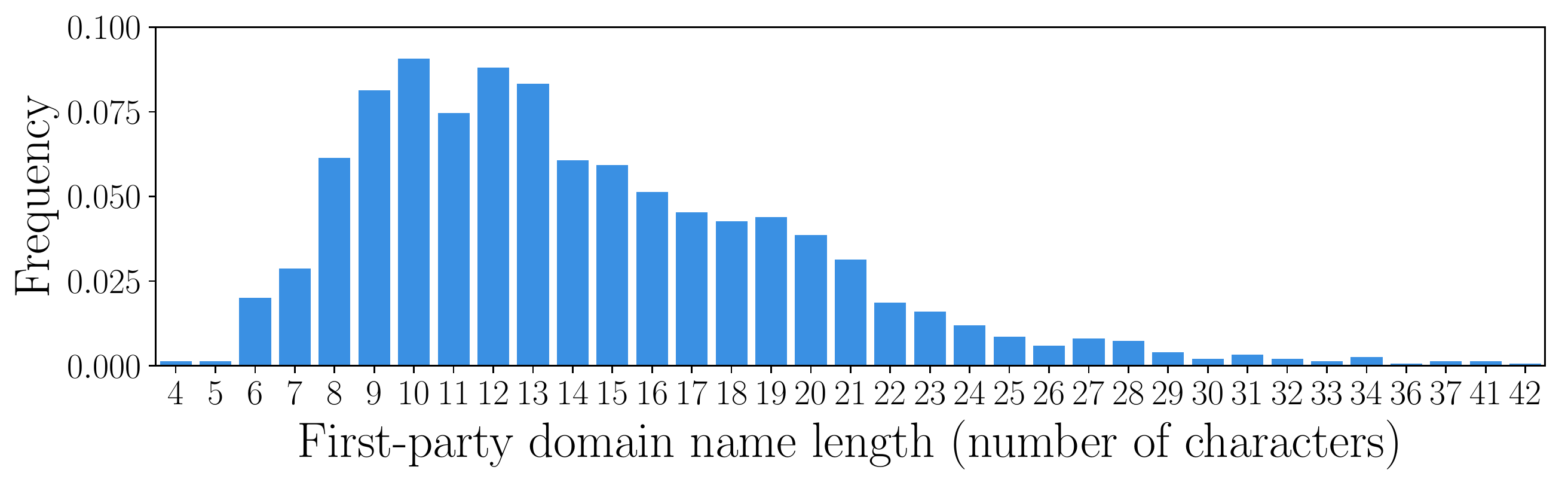}
\end{subfigure}
\begin{subfigure}{0.49\textwidth}
  \centering
  \includegraphics[width=1\linewidth]{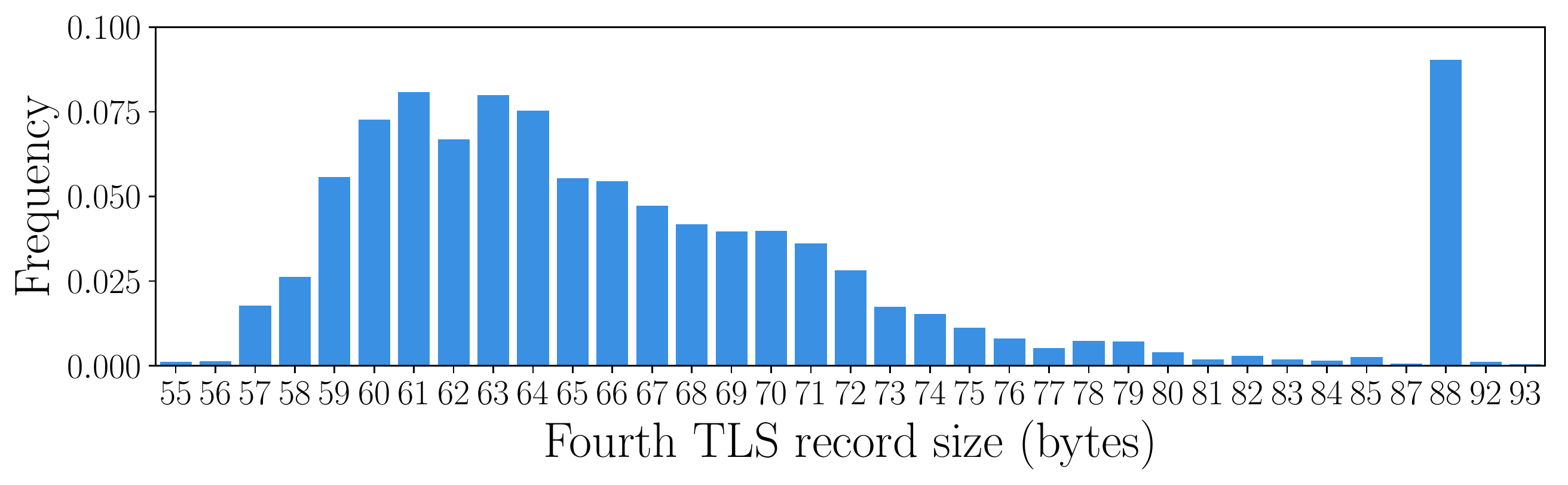}
\end{subfigure}
	\caption{Histograms for domain name length (top) and fourth TLS record length
	(bottom) in the \loca dataset, for 10 samples (normalized over the
	total sum of counts).}
\label{fig:histo_domains}
\vspace{-0.7cm}
\end{figure}

\para{The importance of the fourth packet.}
We hypothesized that the large drop by the fourth packet might be caused by 
one of these records containing the DNS lookup. As these traces do not contain
padding, the domain length would be directly observable in the trace. 
We verify this hypothesis by comparing the frequency of appearance of the domain's and outgoing
fourth record's length (incoming records cannot contain a lookup). 
We discard TLS record sizes corresponding to HTTP2 control messages, e.g., the size ``33'' which
corresponds to HTTP2 acknowledgements and outliers that
occurred 5\% or less times. However, We kept size ``88'' even
though it appears too often, as it could be caused by queries containing 37-characters-long domain names. 

We represent the frequency distributions in Figure~\ref{fig:histo_domains}.
We see that the histogram of the sizes of the fourth TLS record in our
dataset is almost identical to the histogram of domain name lengths, being
the constant difference of 51 bytes between the two histograms
the size of the HTTPS header.
This confirms that the fourth packet often contains the first-party DoH query. 
In some traces the DoH query is sent earlier, explaining the entropy decrease starting
after the second packet.   

\subsection{Censor DNS-blocking strategy}

We study the collateral damage, \ie how many sites are affected
when a censor blocks one site after a traffic-analysis based inference.
We assume that upon decision, the censor uses standard techniques to block 
the connection~\cite{KhattakESSMG16}.

\para{High-confidence blocking.}
To minimize the likelihood of collateral damage, the adversary
could wait to see enough packets for the conditional entropy to 
be lower than one bit. This requires waiting for,
on average, 15\% of the TLS records in the DoH connection
before blocking. 
As those packets include the resolution to the first-party domain,
the client can download  the content served from this domain. Yet, the censor 
can still disrupt access to subsequent queried domains (subdomains and third-parties). 
This is a strategy already used in the wild 
as a stealthy form of censorship~\cite{KhattakESSMG16}.

To avoid this strategy, the clients could create one connection
per domain lookup, thereby mixing connections belonging
to the censored page and others. At the cost of generating more traffic for users and resolvers, this
would force the censor to drop all DoH connections originating from a user's IP or throttle their DoH traffic, causing
more collateral damage.

\para{Block on first DoH query.}
A more aggressive strategy is to drop the DoH connection before the first DoH response
arrives. While this guarantees that the client cannot access any content, not
even \texttt{index.html}, it also results in all domains with same name length being censored.

\begin{figure}[t!] \centering
	\includegraphics[width=0.47\textwidth]{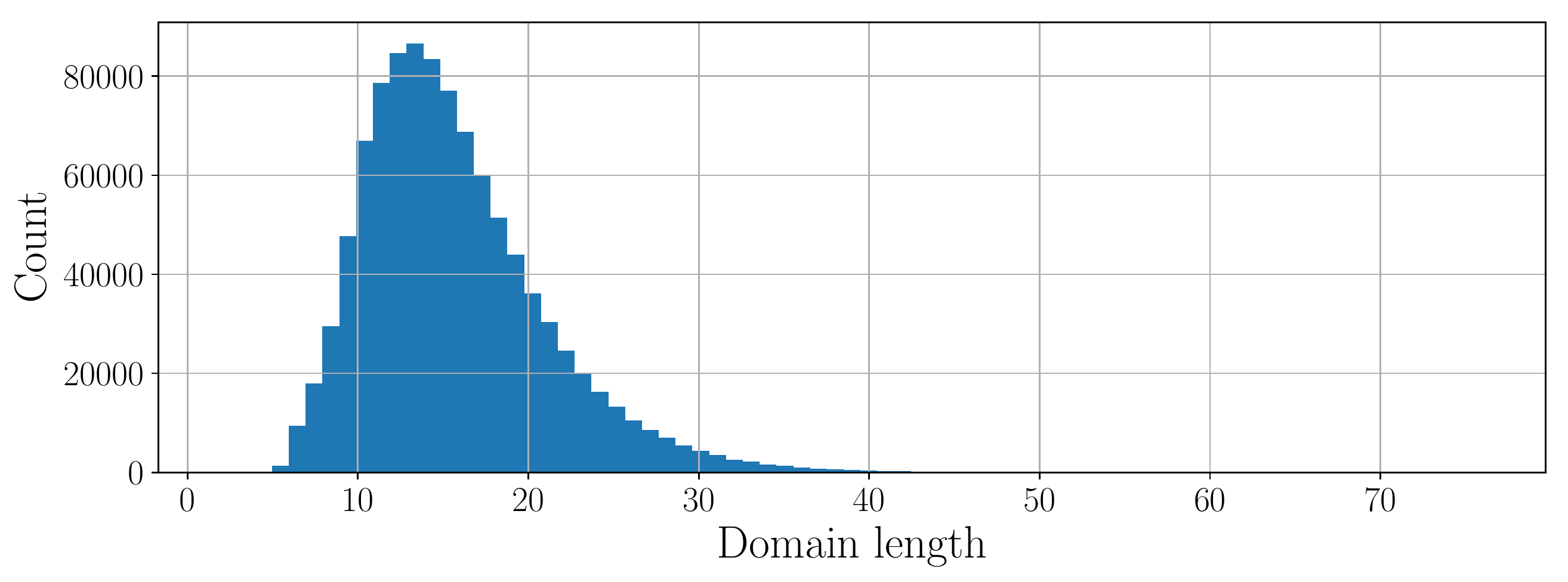}
 	\caption{Distribution of domain length in the Alexa top 1M.}
	\label{fig:alexa1M}
\vspace{-0.7cm}
\end{figure}

In order to understand the collateral damage incurred by domain length based blocking 
relying on the fourth packet, we compare the distribution of domain name lengths in 
the Alexa top 1M ranking (see Fig.~\ref{fig:alexa1M}) with the distribution of domain names
likely to be censored in different countries. For the former 
we take the global ranking as per-country rankings only provide 500 top websites which
are not enough for the purpose of our experiments and, for some countries, the lists are not even available.
We take the test lists provided by Citizen Labs~\cite{TestingLists}. These
lists contain domains that regional experts identify as likely to be censored.
While appearance in these lists does not guarantee that the domains are actually
censored, we believe that they capture potential censor behavior. 

\begin{figure*}[t!] \centering
	\includegraphics[width=0.8\textwidth]{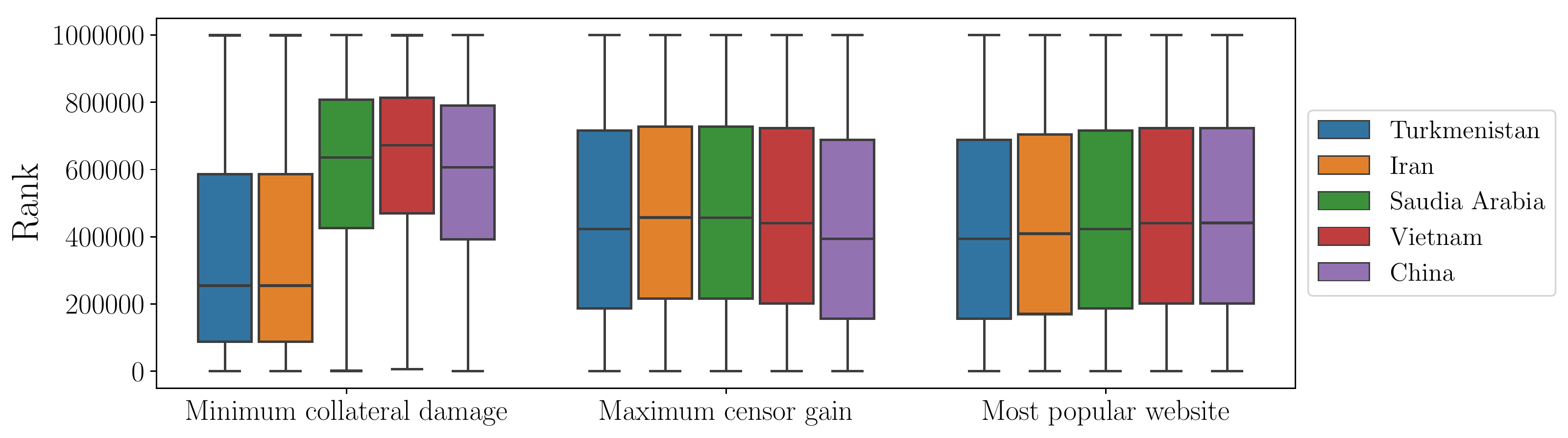}
 	\caption{Collateral damage in terms of ranking for three blocking strategies: minimum collateral damage
 		set size, maximum censor gain, most popular website.}
	\label{fig:ranks}
\vspace{-0.7cm}
\end{figure*}

We take five censorship-prone countries as examples: 
Turkmenistan, Iran, Saudi Arabia, Vietnam, and China.
Since the Alexa list contains only domains, we extract the domains from the URLs
appearing in Citizen Labs' test lists. Table~\ref{tab:censoredinalexa}
shows the total number of domains in each list and the number of those domains that
are not present in the Alexa-1M list. We observe that at most
51\% (China) of the domains appear in the ranking. For the other countries
the ratio is between 21\% and 32\%. This indicates that the
potentially censored domains themselves are mostly not popular.

\begin{table}[t!]
  \centering
  \caption{Number of domains in each censorship test list and their presence in the Alexa top 1M.}
  \resizebox*{\columnwidth}{!}{%
  \begin{tabular}[c]{ l  c  c  c  c  c  c }
   \toprule
      & \textbf{Turkmenistan} & \textbf{Iran} & \textbf{S. Arabia } & \textbf{Vietnam} & \textbf{China}  \\
    \midrule
    	Censored domains & 344 & 877 & 284 & 274 & 191 \\
    	Not in Alexa-1M & 246 & 600 & 219 & 218 & 94 \\
    \bottomrule
  \end{tabular}}
  \label{tab:censoredinalexa}
\end{table}

While in themselves these domains are not popular, as the censor
can only use their length to identify related DoH lookups, there
are two effects. First, there will be some collateral damage: allowed websites 
with the same domain length will also be blocked. 
Second, there will be a gain for the censor: censored websites with 
the same domain length are blocked at the same time.

First, we study the trade-off between collateral damage and censor gain.
The minimum collateral damage is attained when
the censor blocks domains of length 5 (such as \texttt{ft.com}), resulting on 1,318 affected websites.
The maximum damage happens when censoring 
domains of size 10 (such as \texttt{google.com}) which affects other 66,923 websites,
domains of size 11 (such as \texttt{youtube.com}) which affects other 78,603 websites,
domains of size 12 (such as \texttt{facebook.com}) which affects other 84,471 websites,
domains of size 13 (such as \texttt{wikipedia.org}) which affects other 86,597 websites,
and domains of size 14 (such as \texttt{torproject.org}, which affects other 83,493 websites.
This means that, in the worst case, collateral damage is at most 8.6\% of the top 1M list.

To understand the censor gain, we consider Iran as an example. 
The maximum gain is 97, for domain length of 13, \ie when the censor blocks domains of length
13, it blocks 97 domains in the list. At the same time, it 
results in large collateral damage (86,557 websites). The minimum
gain is obtained for domain length 5, which only blocks one website, but
causes small collateral damage (1,317 domains). 
If Iran blocks the most popular domain in the list (\url{google.com}), this results in a collateral damage of 66,887
websites.

The volume of affected websites is of course representative of damage, but it is not
the only factor to take into account. Blocking many non-popular domains that are in
the bottom of the Alexa list is not the same as blocking a few in the top-100. We show in
Figure~\ref{fig:ranks} boxplots representing the distribution of Alexa ranks for three
blocking strategies: minimum collateral damage, maximum censor gain,
and blocking the most popular web. For each country, these strategies require blocking
different domain lengths. Thus, in terms of volume, the damage is different but in the order of
the volumes reported above. We observe that the minimum collateral damage strategy results in different impact depending on the country. 
Although for all of them the volume is much lower than for the other strategies, in Turkmenistan and Iran the median rank
of the blocked websites is much lower than those in the other strategies, indicating that this strategy may have potentially higher impact
than blocking more popular or high-gain domains. 
On the positive side, minimum collateral damage in Saudi Arabia, Vietnam, and China, mostly affects high-ranking websites. Thus,
this strategy may be quite affordable in this country.
The other two strategies, maximum gain and blocking the most popular website, result on a larger number of websites blocked, but their median ranking is high (above 500,000)
and thus can also be considered affordable. 

In conclusion, our study shows that while block on first DoH query is an aggressive strategy,
it can be affordable in terms of collateral damage. More research is needed to fine-tune 
our analysis, e.g., with access to large per-country rankings, or exact lists of blacklisted domains.

\section{Looking ahead}
\label{sec:conclusion}
%!TEX root = report.tex

We have shown that, although it is a great step for privacy, 
encrypting DNS does not completely prevent monitoring or 
censorship. Current padding strategies have great potential to
prevent censorship, but our analysis shows that they
fall short when it comes to stopping resourceful adversaries from monitoring
users' activity on the web.

Our countermeasures analysis hints that the path towards full protection
is to eliminate size information. In fact, the repacketization
in constant-size cells offered by Tor provides the best
practical protection. Tor, however, induces a large overhead both
in the bandwidth required to support onion encryption, as well as
in download time due to the rerouting of packets through the Tor network.

We believe that the most promising approach to protect DoH is to
have clients mimicking the repacketization strategies of Tor, 
without replicating the encryption scheme or re-routing.
This has the potential to improve the trade-off between overhead 
and traffic analysis resistance. A complementary strategy to ease
the achievement of constant-size flows, is to rethink the format
of DNS queries and its headers. Reducing the number of bits required
for the headers would make it easier to fit queries and responses
in one constant-size packet with small overhead.

Besides protection against third party observers, it is important that
the community also considers protection from the resolvers. The current
deployment of DoH, both on the side of resolvers and browsers,
concentrates all lookups in a small number of actors that can observe
the behavior of users. More research in the direction of Oblivious 
DNS~\cite{schmitt2019oblivious} is needed to ensure that no parties can become
main surveillance actors.

\bibliographystyle{unsrt}
\bibliography{bibliography}

\appendix
%!TEX root = report.tex
\subsection{Performance metrics.} \label{sec:metrics}

In this section, we provide an overview of our classifer evaluation metrics.
We use standard metrics to evaluate the performance of our classifier:
\emph{Precision}, \emph{Recall} and \emph{F1-Score}. We compute these metrics
per class, where each class represents a webpage. We compute these metrics
on a class as if it was a ``one vs. all'' binary classification: we
call ``positives'' the samples that belong to that class and ``negatives'' the
samples that belong to the rest of classes. Precision is the ratio
of true positives to the total number of samples that were classified as
positive (true positives and false positives). Recall is the ratio of true positives
to the total number of positives (true positives and false 
negatives). The F1-score is the harmonic mean of precision and
recall. 

%!TEX root = report.tex

\subsection{Estimation of Probabilities}~\label{sec:prob_estimates}

In this section, we explain how we estimated the probabilities for the
entropy analysis in Section~\ref{sec:censorship}.

We define the \emph{anonymity set} of a trace $s$ as a multiset:

$$ A(s) := \{w^{\mbox{m}_s(w)}\},$$

where $\mbox{m}_s(w)$ is the multiplicity of a website $w$ in $A(s)$. The
multiplicity is a function defined as the number of times that trace $s$
occurrs in $w$.%, i.e., $\mbox{m}_s(w) := |\{s_j\in w : s_j = o\}|$.

The probability $\Pr[W=w\mid S_l = o]$ can be worked out using Bayes. For
instance, for website $w$,

\begin{equation}
\label{eq:bayes}
\Pr[W=w\mid S_l=o] = \frac{\Pr[W=w]\Pr[S_l=o\mid W=w]}{\displaystyle\sum_{i=1}^m
\Pr[W=w_i]\Pr[S_l=o\mid W=w_i]}
\end{equation}

We assume the distribution of priors is uniform, i.e., the
probability of observing a website is the same for all websites:
$\Pr[w_i] = \Pr[w_j]\quad\forall i, j$.

We acknowledge that this is an unrealisitc assumption but we provide the
mathematical model to incorporate the priors in case future work has the data
to estimate them.

Assuming uniform priors allows us to simplify the Bayes rule formula since we
can factor out $\Pr[W=w_i]$ in Equation~\ref{eq:bayes}

Regarding the likelihoods of observing the traces given a website, we can use
the traffic trace samples in our dataset as observations to estimate them:

$$\Pr[S_l=o\mid W=w_i] \approx \frac{\mbox{m}_s(w_i)}{k_i}$$

Since we have a large number of samples for all the sites, we can fix the same
sample size for all sites: $k_i = k_j\quad\forall i,j$. A fixed sample size
allows us to factor out $k_i$ in our likelihood estimates and, thus, the
posterior can be estimated simply as:

$$\Pr[W=w\mid S_l=o] \approx \frac{\mbox{m}_s(w)}{\displaystyle\sum_{i=1}^m
\mbox{m}_s(w_i)} = \frac{\mbox{m}_s(w)}{|A(s)|} $$.

That is the multiplicity of website $w$ divided by the size of the $s$'s
anonymity set, which can be computed efficiently for all $w$ and $s$ using
vectorial operations.

\subsection{Extra results on attack robustness}
\label{sec:extra}

\begin{figure}[h]
\centering
\includegraphics[width=1\linewidth]{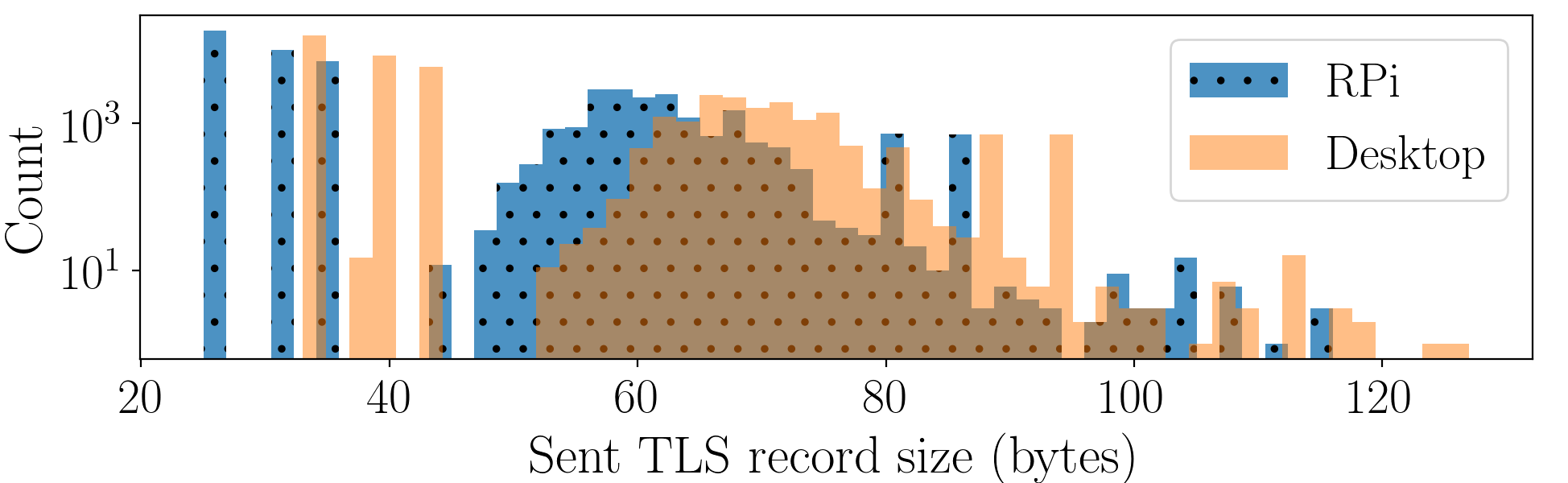}
\caption{Distribution of user's sent TLS record sizes in platform experiment.}
\label{fig:histplatform}
\end{figure}

In this section, we provide additional information on our experiment measuring
the influence of end user's platform (Section~\ref{sec:infra}). Figure~\ref{fig:histplatform}
shows the difference in TLS record sizes for both platforms -- the sizes follow a similar 
distribution, with a shift. Table~\ref{tab:platformimp} shows the improvement in
performance we get when removing this shift. 
 
\begin{table}[htbp]
  \centering
  \caption{Improvement in cross platform performance when removing the shift 
  (standard deviation less than 1\%).}
  \begin{tabular}[c]{  c c  c  c  c  }
    \toprule
    \textbf{Train} & \textbf{Test} &  \textbf{Precision} & \textbf{Recall} & \textbf{F-score}\\
    \midrule
     \desktop  & \rpi & $0.630$ & $0.654$ & $0.614$ \\
      \rpi & \desktop & $0.552$ & $0.574$ & $0.535$ \\
    \bottomrule
  \end{tabular}
  \label{tab:platformimp}
\end{table}

\subsection{Survivors and Easy Preys}\label{app:survivors}

In this section, we show results from our analysis of survivors and
easy preys, as discussed in Section~\ref{sec:robustness}. We show
the top-10 webpages with highest-mean and lowest-variance (Table~\ref{tab:doomed}), 
lowest-mean and lowest-variance (Table~\ref{tab:survivors}), 
and highest-variance F1-score (Table~\ref{tab:high_variance}).

\begin{table}[h]
\centering
      \caption{Top-10 with highest-mean and lowest-variance F1-Score}\label{tab:doomed}
    \begin{tabular}{rrrl}
    \toprule
    Alexa Rank &  Mean F1-Score &  Stdev F1-Score &                    Domain name \\
    \midrule
     777 &              0.95 &             0.08 &     militaryfamilygiftmarket.com \\
     985 &              0.95 &             0.08 &                       myffpc.com \\
     874 &              0.95 &             0.08 &          montrealhealthygirl.com \\
     712 &              0.95 &             0.08 &                mersea.restaurant \\
    1496 &              0.95 &             0.08 &              samantha-wilson.com \\
    1325 &              0.95 &             0.08 &          nadskofija-ljubljana.si \\
     736 &              0.95 &             0.08 &               michaelnewnham.com \\
     852 &              0.95 &             0.08 &            mollysatthemarket.net \\
     758 &              0.95 &             0.08 &                midwestdiesel.com \\
    1469 &              0.95 &             0.08 &  reclaimedbricktiles.blogspot.si \\
    \bottomrule
    \end{tabular}

%\bigskip

\centering
      \caption{Top-10 sites with lowest-mean and lowest-variance F1-Score}\label{tab:survivors}
    \begin{tabular}{rrrl}
    \toprule
    Alexa Rank &  Mean F1-Score &  Stdev F1-Score &          Domain name \\
    \midrule
     822 &              0.11 &             0.10 &          mjtraders.com \\
    1464 &              0.11 &             0.08 &        ravenfamily.org \\
     853 &              0.14 &             0.09 &  moloneyhousedoolin.ie \\
     978 &              0.14 &             0.17 &   mydeliverydoctor.com \\
     999 &              0.17 &             0.10 &  myofascialrelease.com \\
     826 &              0.17 &             0.11 &             mm-bbs.org \\
    1128 &              0.17 &             0.10 &          inetgiant.com \\
     889 &              0.18 &             0.14 &           motorize.com \\
     791 &              0.18 &             0.15 &        mindshatter.com \\
    1193 &              0.20 &             0.14 &   knjiznica-velenje.si \\
    \bottomrule
    \end{tabular}

\centering
      \caption{Top-10 sites with highest-variance F1-Score}\label{tab:high_variance}
    \begin{tabular}{rrrl}
    \toprule
    Alexa Rank &  Mean F1-Score &  Stdev F1-Score &                      Domain name \\
    \midrule
    1136 &              0.43 &             0.53 &    intothemysticseasons.tumblr.com \\
     782 &              0.43 &             0.53 &                   milliesdiner.com \\
     766 &              0.43 &             0.53 &      mikaelson-imagines.tumblr.com \\
    1151 &              0.43 &             0.53 &  japanese-porn-guidecom.tumblr.com \\
     891 &              0.42 &             0.52 &        motorstylegarage.tumblr.com \\
     909 &              0.42 &             0.52 &          mr-kyles-sluts.tumblr.com \\
     918 &              0.44 &             0.52 &       mrsnatasharomanov.tumblr.com \\
    1267 &              0.52 &             0.49 &              meander-the-world.com \\
     238 &              0.48 &             0.49 &                     caijing.com.cn \\
     186 &              0.48 &             0.48 &                           etsy.com \\
    \bottomrule
    \end{tabular}

\end{table}

\subsection{Confusion Graphs}

We have used \emph{confusion graphs} to understand the errors of the classifier.
Confusion graphs are the graph representation of confusion matrices. They allow to
easily visualize large confusion matrices by representing misclassifications as
directed graphs. Confusion graphs have been used in website
fingerprinting~\cite{OverdorfJAGD17} and other classification tasks to
understand classifier error~\cite{YoshidaB16}.

\begin{figure}[h]
  \centering
    \includegraphics[width=0.4\textwidth]{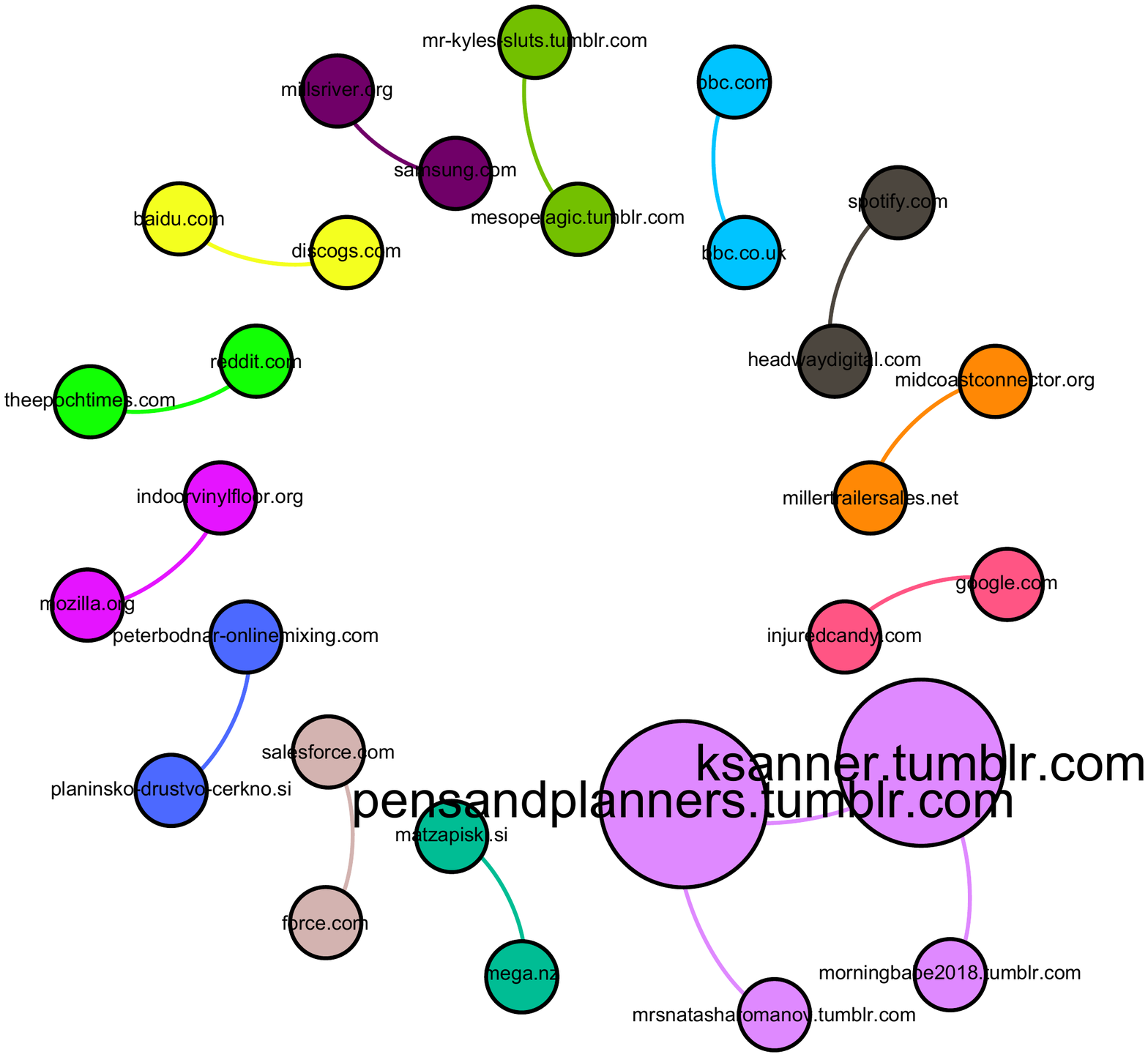}
  \caption{Confusion graph for the misclassifications in \loca that happen in more than one fold of the cross-validation and have different domain name length. We observe domains that belong to the same CDN (e.g., tumblr) or belong to the same entity (e.g., BBC, Salesforce). For others, however, the cause of the misclassification remains an open question.}
  \label{graph:outliers_wo_one_time_diff_len}
\end{figure}

\begin{figure}[h]
  \centering
    \includegraphics[width=0.4\textwidth]{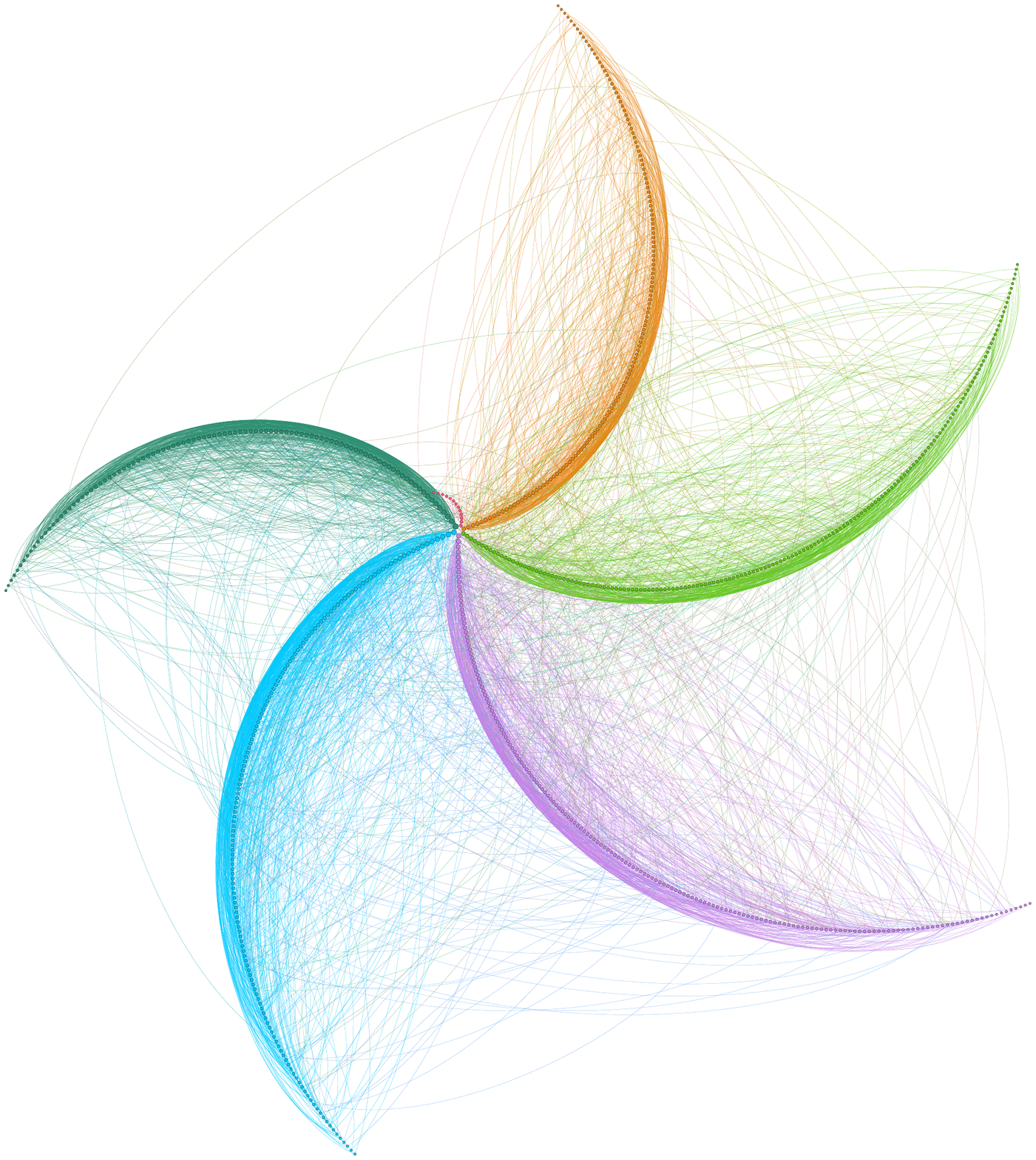}
  \caption{Confusion graph for all Tor misclassifications. We did not plot the
labels to remove clutter. We observe that domains in one ``petal'' of the graph
tend to classify between each other.}
  \label{graph:tor_misclassif_multiple}
\end{figure}

\begin{figure}[ht]
  \centering
    \includegraphics[scale=0.8,angle=90]{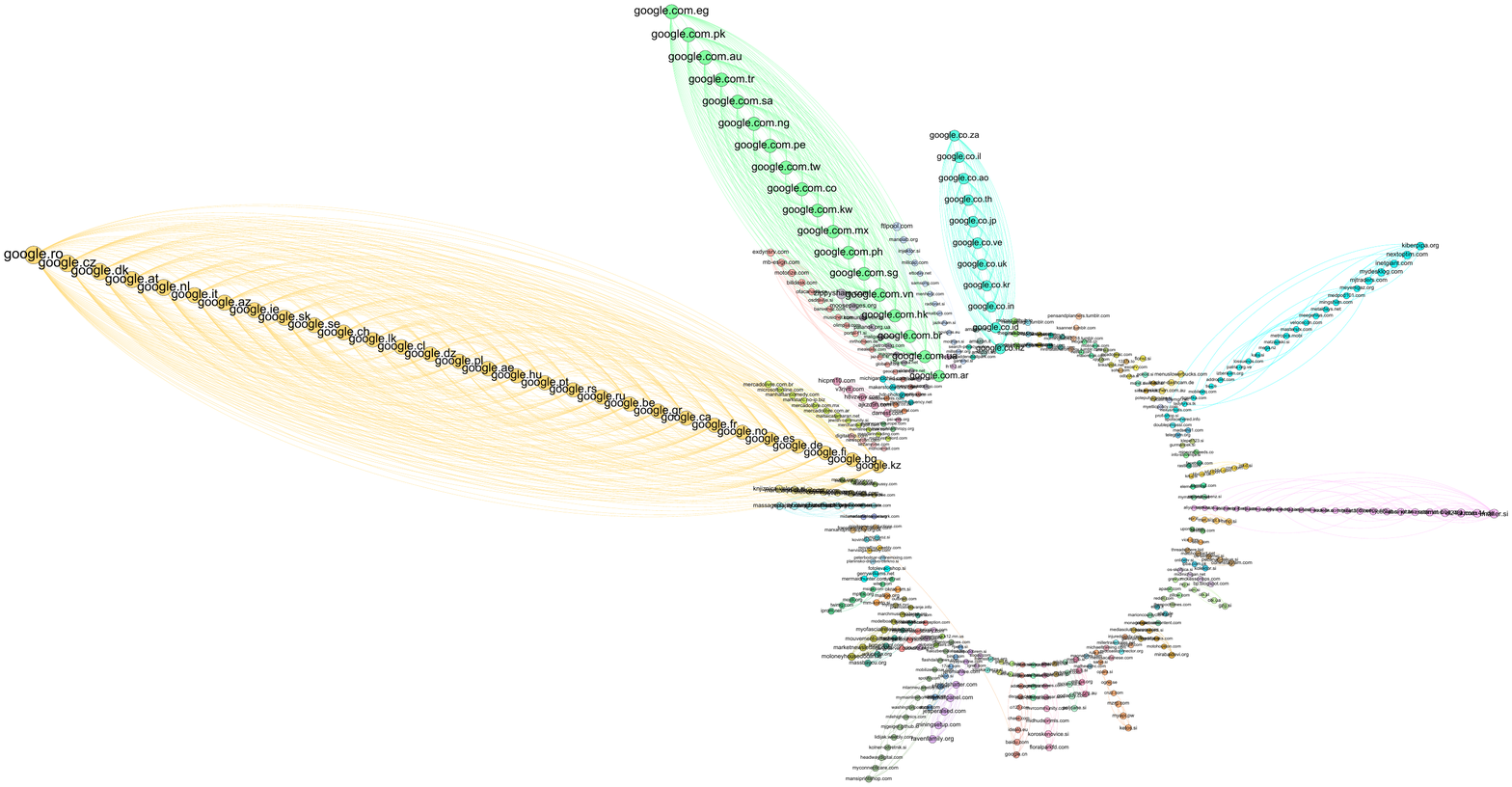}
  \caption{Confusion graph for all misclassifications in \loca. We observe clusters of domains such as Google and clusters of domains that have the same name length. Interestingly, the
only inter-cluster edge we observe is between one of the Google clusters and a cluster that mostly contains Chinese domains.}
  \label{graph:outliers_wo_one_time}
\end{figure}

Figures~\ref{graph:outliers_wo_one_time_diff_len},~\ref{graph:tor_misclassif_multiple}~and~\ref{graph:outliers_wo_one_time}
show the classification errors in the form of confusion graphs for some of the
experiments presented in Sections~\ref{sec:fingerprintability} and ~\ref{sec:countermeasures}. 
 The graphs were
drawn using Gephi, a software for graph manipulation and visualization.  Nodes
in the graph are domains and edges represent misclassifications between
domains. The edge source is the true label of the sample and the destination is
the domain that the classifier confused it with. The direction of the edge is
encoded clockwise in the curvature of the edge.  Node size is proportional to
the node's degree and nodes are colored according to the community they belong
to, which is determined by the Lovain community detection
algorithm~\cite{BlondelGLL08}.

\begin{figure}[h]
  \centering
    \includegraphics[width=0.4\textwidth]{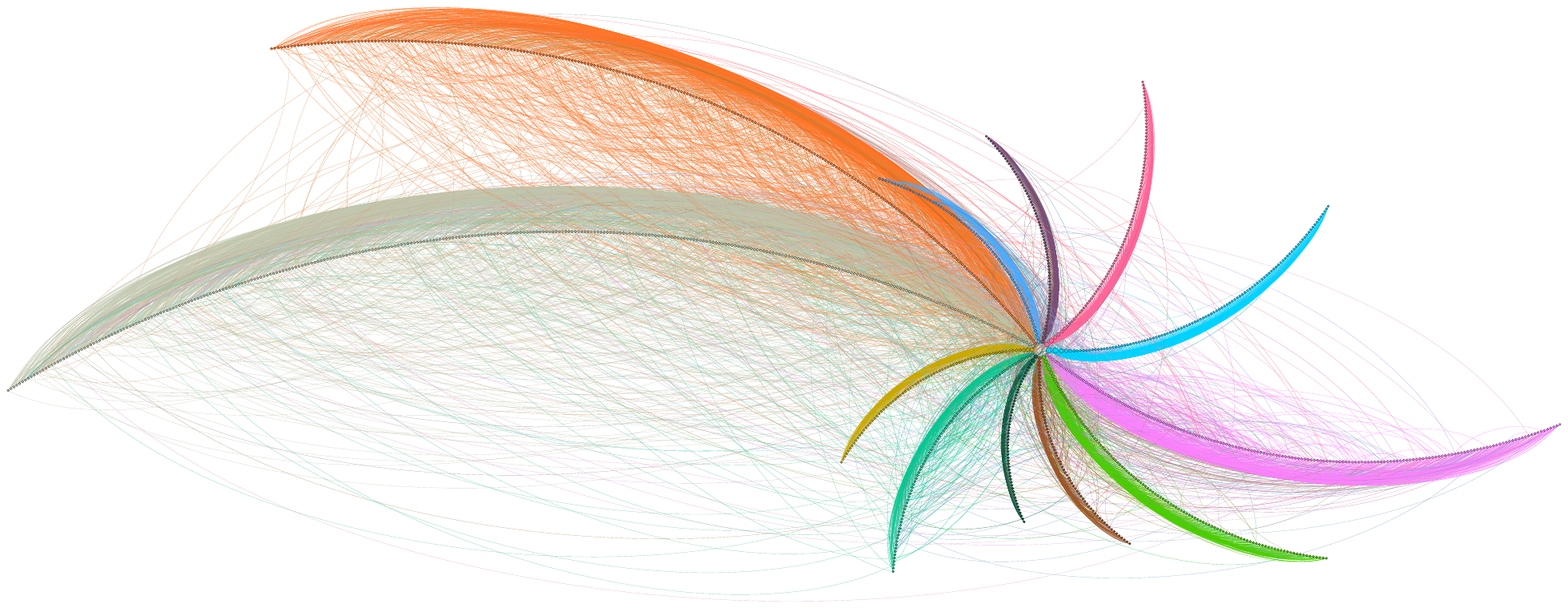}
  \caption{Confusion graph for all perfect padding misclassifications. We observe
  a larger number of petals as compared to DNS-over-Tor. Domains in one petal of the graph
tend to classify between each other.}
  \label{graph:const_padding_star}
\end{figure}

%\newpage
\onecolumn
\newpage

\newpage

\end{document}